\documentclass[sigconf]{acmart}

\setcopyright{none} 
\acmConference[]{}{}

\usepackage{subcaption}
\usepackage{float}
\usepackage{booktabs}
\usepackage{tikz}
\usetikzlibrary{shapes.geometric, arrows}
\usepackage{url}
\usepackage{epigraph}

\newcommand{\ie}{{i.e.,~}}
\newcommand{\eg}{{e.g.,~}}

\newcommand{\xpr}{{X_{priv}}}
\newcommand{\txpr}{{\tilde{X}_{priv}}}
\newcommand{\xpu}{{X_{pub}}}
\newcommand{\ypu}{{Y_{pub}}}
\newcommand\myeq{\mkern1.5mu{=}\mkern1.5mu}
\newcommand{\f}{f}
\newcommand{\fc}{C_{att}}
\newcommand{\s}{s}
\newcommand{\tf}{\tilde{f}}
\newcommand{\tfi}{\tf^{-1}}
\newcommand{\Z}{\tilde{\mathbf{Z}}}

\newcommand{\X}{\mathbf{X}}
\newcommand{\loss}{\mathcal{L}}
\newcommand{\TT}[1]{``\textit{#1}''}
\newcommand{\new}[1]{\textcolor{black}{#1}}

\newcommand{\mb}[1]{\textcolor{black}{#1}}

\usepackage{balance}

\newcommand{\todobox}[3]{%
	\colorbox{#1}{\textcolor{white}{\sffamily\bfseries\scriptsize #2}}%
	~\textcolor{red}{#3} %
	\textcolor{#1}{$\triangleleft$}%
}
\newcommand{\todo}[1]{\todobox{red}{TODO}{#1}}

\usepackage[export]{adjustbox}
\usepackage[most]{tcolorbox}
\usepackage{epstopdf}
\usepackage[linesnumbered, noend, ruled,vlined]{algorithm2e}

\SetCommentSty{mycommfont}

\usepackage[bottom]{footmisc}

\begin{document}
\title{Unleashing the Tiger: Inference Attacks on Split Learning}

\author{Dario Pasquini}
\affiliation{%
	\institution{EPFL}
	 \city{Lausanne}
	 \country{Switzerland}
}
\email{dario.pasquini@epfl.ch} 
\author{Giuseppe Ateniese} \affiliation{%
	\institution{George Mason University}
	\city{Fairfax}
	\state{Virginia}
	\country{USA}
}
\email{ateniese@gmu.edu} 

\author{Massimo Bernaschi} 
\affiliation{%
	\institution{Institute of Applied Computing, CNR}
    \city{Rome}
	\country{Italy}
}
\email{massimo.bernaschi@cnr.it} 

\begin{abstract}
We investigate the security of \textit{split learning}---a novel collaborative machine learning framework that enables peak performance by requiring minimal resource consumption. In the present paper, we expose vulnerabilities of the protocol and demonstrate its inherent insecurity by introducing general attack strategies targeting the reconstruction of clients' private training sets. More prominently, we show that a malicious server can actively hijack the learning process of the distributed model and bring it into an insecure state that enables inference attacks on clients' data. We implement different adaptations of the attack and test them on various datasets as well as within realistic threat scenarios. We demonstrate that our attack can overcome recently proposed defensive techniques aimed at enhancing the security of the split learning protocol. Finally, we also illustrate the protocol's insecurity against malicious clients by extending previously devised attacks for Federated Learning.
\end{abstract}

\maketitle
\section{Introduction}
\epigraph{\textit{Once the cattle have been split up, then the tiger strikes.}}{A Myanma proverb}
Deep learning requires massive data sets and computational power. State-of-the-art neural networks may contain millions or billions~\cite{gpt3} of free parameters and necessitate representative training sets. Unfortunately, collecting suitable data sets is difficult or sometimes impossible. Entities and organizations may not be willing to share their internal data for fear of releasing sensitive information. For instance, telecommunication companies would benefit extraordinarily from deep learning techniques but do not wish to release customer data to their competitors. Similarly, medical institutions cannot share information because privacy laws and regulations shelter patient data.

Secure data sharing and learning can only be achieved via cryptographic techniques, such as homomorphic encryption or secure multi-party computation. However, the combination of cryptography and deep learning algorithms yields expensive protocols. 
An alternative approach, with mixed results, is distributed/decentralized machine learning, where different parties cooperate to learn a shared model. In this paradigm, training sets are never shared directly. In federated learning~\cite{federated0, federated1, federated2}, for example, users train a shared neural network on their respective local training sets and provide only model parameters to others. The expectation is that by sharing certain model parameters, possibly \TT{scrambled}~\cite{dp}, the actual training instances remain hidden and inscrutable. Unfortunately, in~\cite{gan_attack}, it was shown that an adversary could infer meaningful information on training instances by observing how shared model parameters evolve over time.
\par

Split learning is another emerging solution that is gaining substantial interest in academia and industry. \textbf{In the last few years, a growing body of empirical studies~\cite{splitnn2, splitnn3, splitnn4, splitnn_what, splitnn_selfsurvey, split_iot,lim2020incentive,split_wave, splitmc,thapa2020advancements}, model extensions~\cite{splitfed, split_fed2, spliteasy, splitnnsec, splitnn8, split_para, splitrnn, sharma2019expertmatcher, splitvert2}, and 
events~\cite{CVPR_tut,SLDML} attested to the effectiveness, efficiency, and relevance of the split learning framework.} At the same time, split-learning-based solutions have been implemented and adopted in commercial as well as open-source applications~\cite{split_com1, split_com0}. Several start-ups, which are receiving much attention, are currently relying on the split learning framework to develop efficient collaborative learning protocols and train deep models on real-world data.
\par

The success of split learning is primarily due to its practical properties. Indeed, compared with other approaches such as federated learning, split learning requires consistently fewer resources from the participating clients, enabling lightweight and scalable distributed training solutions. However, while the practical properties of split learning have been exhaustively validated~\cite{splitnn_selfsurvey, splitnn4}, little effort has been spent investigating the security of this machine learning framework.
\par

In this paper, \textbf{we carry out the first, in-depth, security analysis of split learning and draw attention to its inherent insecurity.} We demonstrate that the assumptions on which the security of split learning is based are fundamentally flawed, and a \mb{motivated} adversary can easily subvert the defenses of the training framework. In particular, we implement a general attack strategy that allows a malicious server to recover private training instances during the distributed training. In the attack, the server hijacks the model's learning processes and drives them to an insecure state that can be exploited for inference attacks.
 In the process, the attacker does not need to know any portion of the client's private training sets or the client's architecture. The attack is domain-independent and can be seamlessly applied to various split learning variants~\cite{splitfed, split_fed2}. We call this general attack: the \textbf{feature-space hijacking attack} (FSHA) and introduce several adaptations of it. We test the proposed attacks on different datasets and demonstrate their applicability under realistic threat scenarios such as data-bounded adversaries.
 \par
 
Furthermore, we show that client-side attacks that have been previously devised on federated learning \mb{settings} remain effective within the split learning framework. In particular, we adapt and extend the inference attack proposed in~\cite{gan_attack} to make it work in split learning. Our attack demonstrates how a malicious client can recover suitable approximations of private training instances of other honest clients participating in the distributed training. Eventually, this result confirms the insecurity of split learning also against client-side attacks.
\par

 The contributions of the present paper can be then summarized as follows:
 \begin{itemize}
	\item We demonstrate the insecurity of split learning against a \textbf{malicious server} by devising a novel and general attack framework. Such a framework permits an attacker to \textbf{(1)}~recover precise reconstructions of individual clients' training instances as well as \textbf{(2)}~perform property inference attacks~\cite{hackingsmartm} for arbitrary attributes. Additionally, we show that the proposed attacks can circumvent defensive techniques devised for split learning~\cite{splitnnsecex, splitnnsec}.
 	\item We demonstrate the insecurity of split learning against a \textbf{malicious client} by adapting and extending previously proposed techniques targeting federated learning~\cite{gan_attack}. The attack permits a malicious client to recover prototypical examples of honest clients' private instances.
 \end{itemize}
To make our results reproducible, we made our code available\footnote{\url{https://github.com/pasquini-dario/SplitNN_FSHA}}.
%
\paragraph{Overview}
The paper starts by surveying distributed machine learning frameworks in Section~\ref{sec:cml}. Section~\ref{sec:fsha} follows by introducing and validating our main contribution---the feature-space hijacking attack framework. Then, Section~\ref{sec:def} covers the applicability of existing defensive mechanisms within the split learning framework. In Section~\ref{sec:cattack}, we analyze the security of split learning against malicious clients. Section~\ref{sec:conc} concludes the paper, with Appendices containing additional material. In the paper, background and analysis of previous works are provided, when necessary, within the respective sections.

\section{Distributed Machine Learning}
\label{sec:cml}
Distributed (also collaborative~\cite{CCS15_pp}) machine learning allows a set of remote clients $\mathbf{Cs}=\{c_1, \dots,c_n\}$ to train a shared model $F$. Each client $c_i$ participates in the training protocol by providing a set of training instances $\xpr_i$. This set is private and must not be directly shared among the parties running the protocol. For instance, hospitals cannot share patients' data with external entities due to regulations such as HIPAA \cite{PMID:12686707}.
\par

In this section, we focus on distributed machine learning solutions for deep learning models. In particular, we describe: \textbf{(1)}~\textit{Federated learning}~\cite{federated0, federated1, federated2} which is a well-established learning protocol\footnote{In the paper, we use the term \TT{federated learning} to refer to the framework proposed in~\cite{federated0, federated1, federated2}  rather than the \TT{federated learning task} .} and \textbf{(2)}~\textit{split learning}~\cite{splitnn, splitnn2, splitnn3} a recently proposed approach that is gaining momentum due to its attractive practical properties. 

\begin{figure*}[t]
	\centering
	\begin{subfigure}{.4\textwidth}
		\resizebox{1\textwidth}{!}{%
		\begin{tikzpicture}
		\tikzstyle{data} = [rectangle, rounded corners, minimum width=1cm, minimum height=1cm,text centered, draw=black, fill=white!30]
		\tikzstyle{net} = [rectangle, minimum width=2cm, minimum height=1cm,text centered, draw=black, fill=cyan!30]
		\tikzstyle{arrow} = [thick,->,>=stealth]
		\tikzstyle{gradient}  = [thick, red,->,>=stealth, opacity=.7]
		
		%
		%
		%
		
		\node (sa)[rectangle, rounded corners, minimum width=5cm, minimum height=3cm,text centered, draw=black, fill=white, opacity=.3, yshift=0, xshift=2.5cm, label=above: Server:] {};
		
		\node (0) [data, yshift=0, xshift=-4cm] {$\xpr$};
		\node (f) [net, right of=0, xshift=1cm] {$\f$};
		\node (f2) [net, right of=0, xshift=1cm, yshift=-.1cm, opacity=0] {$\f$};
		
		\node (s) [net, right of=f, xshift=3cm] {$\s$};
		\node (s2) [net, right of=f, xshift=3cm, yshift=-.1cm, opacity=0] {$\s$};
		\node (loss) [right of=s, xshift=1cm] {$\loss_{\f, \s}$};
		
		\draw [gradient] (s2.west) to [out=-150,in=-30] (f2.east);
		
		\draw [arrow] (0) -- (f);
		\draw [arrow] (f) -- (s);
		\draw [arrow] (s) -- (loss);


		\end{tikzpicture}
	}
	\caption{Split learning.}
	\label{fig:split_a}
	\end{subfigure}\hspace{2cm}\begin{subfigure}{.4\textwidth}
\resizebox{1\textwidth}{!}{%
	\begin{tikzpicture}
\tikzstyle{data} = [rectangle, rounded corners, minimum width=1cm, minimum height=1cm,text centered, draw=black, fill=white!30]
\tikzstyle{net} = [rectangle, minimum width=2cm, minimum height=1cm,text centered, draw=black, fill=cyan!30]
\tikzstyle{arrow} = [thick,->,>=stealth]
\tikzstyle{gradient}  = [thick, red,->,>=stealth, opacity=.7]

%
%
%

\node (sa)[rectangle, rounded corners, minimum width=4.5cm, minimum height=3cm,text centered, draw=black, fill=white, opacity=.3, yshift=0, xshift=2cm, label=above: Server:] {};

\node (0) [data, yshift=0, xshift=-4cm, yshift=.5cm] {$\xpr$};
\node (f) [net, right of=0, xshift=1cm] {$\f$};
\node (f2) [net, right of=0, xshift=1cm, yshift=-.1cm, opacity=0] {$\f$};
\node (ff) [net, below of=f, yshift=-.3cm] {$\f^{'}$};
\node (ff2) [net, below of=f, yshift=-.5cm, opacity=0] {$\f^{'}$};

\node (loss) [left of=ff, xshift=-1cm] {$\loss_{\f, \f',\s}$};

\node (s) [net, right of=f, xshift=3cm] {$\s$};
\node (s2) [net, right of=f, xshift=3cm, yshift=-.1cm, opacity=0] {$\s$};
\node (s3) [net, right of=f, xshift=3cm, yshift=--.15cm, opacity=0] {$\s$};

\draw [gradient] (s2.west) to [out=-150,in=-30] (f2.east);
\draw [gradient] (ff2.east) to [out=-20,in=-45]  (s3.east);

\draw [arrow] (0) -- (f);
\draw [arrow] (f) -- (s);
\draw [arrow] (ff) -- (loss);
\draw [arrow] (s.east) to [out=-70,in=0]  (ff.east);

	
	\end{tikzpicture}
}
	\caption{Split learning with labels protection.}
	\label{fig:split_b}
\end{subfigure}
\caption{Two variations of split learning. Black arrows depict the activation propagation of the participating neural networks, whereas red arrows depict the gradient that \mb{follows} after the forward pass.}
\label{fig:split}
\end{figure*}
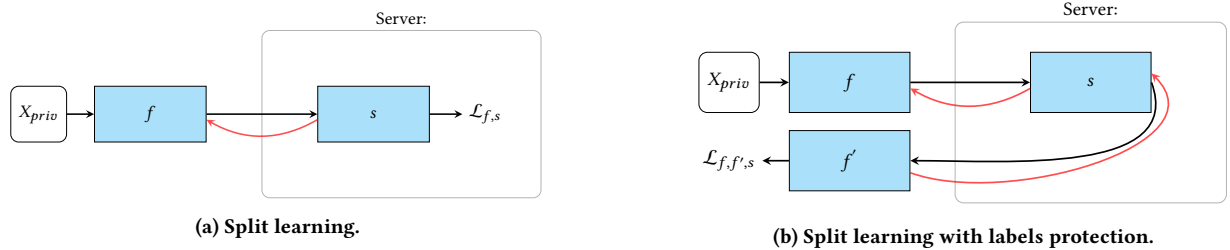

\subsection{Federated Learning}
Federated learning~\cite{federated0, federated1, federated2} allows for distributed training of a deep neural model by aggregating and synchronizing local parameter adjustments among groups of remote clients. In the most straightforward setup, the protocol is orchestrated by a central server that manages clients' training rounds and maintains a master copy of the trained model.
\par

In the initial setup phase, the parties choose a training task and define a machine learning model. The latter is initialized and hosted by the server that makes it available to all remote clients. At each training step, each client downloads the model from the server and locally applies one or more iterations of standard Stochastic Gradient Descent (SGD) using its private training set. After the local training is done, clients send the accumulated gradients (or weights) to the server.\footnote{This process may differ in practice as there are several implementations of federated learning.} The server aggregates these changes into a single training signal applied to the hosted model parameters, completing a global training iteration. Once the server's network is updated, the clients download the new state of the model and repeat the protocol till a stop condition is reached.
\par

At each iteration in federated learning, clients exchange an amount of data with the server that is linear in the number of parameters of the network. For large models, this becomes unsustainable and may limit the applicability of the approach. Several improvements to the framework have been proposed to address this problem~\cite{opt_fed0, opt_fed1}.

\subsubsection{On the security of Federated Learning}
Clients share only gradients/weights induced by the local training steps. The intuition behind federated learning is that local data is safe because it is never directly shared with the server or other clients. Additionally, gradients collected by the server can be further protected through a secure aggregation protocol. The aim is to hinder inference attacks by the server that cannot distinguish clients' individual gradients.
\par

 In federated learning, all the parties have equal access to the trained network. Thus, the server and the clients know the architecture of the network as well as its weights during the various training steps.
\par

Under suitable assumptions, different attacks on federate learning were shown feasible. The first and most prominent is an active attack ~\cite{gan_attack} that allows a malicious client to infer relevant information on training sets of other honest clients by manipulating the learning process. Additionally, the gradients received from the server can be inverted~\cite{DLG}. Other attacks include backdoor injection and poisoning~\cite{fed_poison0, fedbackdoor, fed_poison}. Accordingly, variants of federated learning have been proposed to reduce the effectiveness of those attacks~\cite{fed_poison_defense, fed_def, fed_def1, fed_def2}. They alleviate but do not solve the problems.
\subsection{Split Learning}
Split learning~\cite{splitnn, splitnn2, splitnn3} enables distributed learning by partitioning a neural network in consecutive chunks of layers among various parties; typically, a set of clients and a server. 
In the protocol, the clients aim at learning a shared deep neural network by securely combining their private training sets. The server manages this process and guides the network's training, bearing most of the required computational cost. 
\par

In split learning, training is performed through a vertically distributed back-propagation~\cite{dist} that requires clients to share \textit{only} intermediate network's outputs (referred to as \textit{smashed data}); rather than the raw, private training instances. This mechanism is sketched in Figure~\ref{fig:split}.
In the minimal setup (\ie Figure~\ref{fig:split_a}), a client owns the first $n$ layers $\f$ of the model, whereas the server maintains the remaining neural network $\s$ \ie $F = \s(\f(\cdot))$. Here, the model's architecture and hyper-parameters are decided by the set of clients before the training phase. In particular, they agree on a suitable partition of the deep learning model and send the necessary information to the server.\textbf{ The server has no decisional power and ignores the initial split $\f$.}
\par

At each training iteration, a client sends the output of the initial layers for a batch of private data $\xpr$ (\ie $\f(\xpr)$) to the server. The server propagates this remote activation through the layers~$\s$ and computes the loss. Then, a gradient-descent-based optimization is locally applied to~$\s$. To complete the round, the server sends the gradient up to the input layer of $\s$ to the client that continues the back-propagation locally on $\f$.
\par

In the case of supervised loss functions, the protocol requires the client to share the labels with the server. To avoid that, split learning can be reformulated to support loss function computation on the client-side (Figure~\ref{fig:split_b}). Here, the activation of the last layer of $\s$ is sent to the client that computes the loss function\footnote{The client can also apply additional layers before the loss computation.}, sending the gradient back to the server that continues the back-propagation as in the previous protocol.
\par

Split learning supports the training of multiple clients by implementing a sequential turn-based training protocol. Here, clients are placed in a circular list and interact with the server in a round-robin fashion. On each turn, a client performs one or more iterations of the distributed back-propagation (\ie Figure~\ref{fig:split}) by locally modifying the weights of $\f$. Then, the client sends the new $\f$ to the next client that repeats the procedure. As stated in~\cite{splitnn}, the training process, for suitable assumptions, is functionally equivalent to the one induced by the standard, centralized training procedure. That is, clients converge to the same network that they would have achieved by training a model on the aggregated training sets.
\par

To overcome the sequential nature of the training process, extensions of split learning have been proposed~\cite{splitfed, split_fed2, split_para}. More prominently, in \cite{splitfed}, split learning is combined with federated learning (\ie \textit{splitfed learning}) to yield a more scalable training protocol. Here, the server handles the forward signal of the clients' network in parallel (without aggregating them) and updates the weights of~$\s$. The clients receive the gradient signals and update their local models in parallel. Then, they perform federated learning to converge to a global $\f$ before the next iteration of split learning. This process requires an additional server that is different from the one hosting~$\s$.\footnote{Otherwise, a single server would access both $\f$ and $\s$, violating the security of the protocol.}
\par

Split learning gained particular interest due to its efficiency and simplicity. Namely, it reduces the required bandwidth significantly when compared with other approaches such as federated learning~\cite{splitnn4, splitnn_selfsurvey}. Certainly, for large neural networks, intermediate activation for a layer is consistently more compact than the network's gradients or weights for the full network. Furthermore, the computational burden for the clients is smaller than the one caused by federated learning. Indeed, clients perform forward/backward propagation on a small portion of the network rather than on the whole. This allows split learning to be successfully applied \mb{to} the Internet of Things (IoT) and edge-device machine learning settings~\cite{split_iot,lim2020incentive,split_wave}.
\subsubsection{On the security of Split learning}
\label{sec:splitnnsec}
Split learning has been proposed as a privacy-preserving implementation of collaborative learning~\cite{splitnn, splitnn2, splitnn3, splitnnsec, splitnn_what}. 
In split learning, users' data privacy relies on the fact that raw training instances are never shared; only \TT{\textit{smashed data}} induced from those instances are sent to the server.
\par

The main advantage of split learning in terms of security is that it can hide information about the model's architecture and hyper-parameters. Namely, the server performs its task ignoring the architecture of $\f$ or its weights. \textbf{As assumed in previous works~\cite{splitnn, splitnn2, splitnn3, splitnn_what}, this split is designed to protect the intellectual property of the shared model and to reduce the risk of inference attacks perpetrated by a malicious server.} \new{As a matter of fact, in this setup, the server cannot execute a standard model inversion attack~\cite{model_inv, bah} since it does not have access to the clients' network and cannot make blackbox queries to it.}
\par

We will show that these assumptions are false, and the split learning framework presents several vulnerabilities that allow an attacker to subvert the training protocol and recover clients' training instances. The most pervasive vulnerability of the framework is the server's entrusted ability to control the learning process of the clients' network. A malicious server can guide $\f$ towards functional states that can be easily exploited to recover $\xpr$ data from $\f(\xpr)$. The main issue is that a neural network is a differentiable, smooth function that is naturally predisposed to be functionally inverted. 
 There is no much that can be achieved by splitting it other than a form of \textbf{\textit{security through obscurity}}, which is notoriously inadequate \mb{since it} gives only a false sense of security.
 \par
In the next section, we empirically demonstrate how a malicious server can exploit the split learning framework's inherent shortcomings to disclose clients' private training sets completely.
Furthermore, in Section~\ref{sec:cattack}, we demonstrate that split learning does not protect honest clients from malicious ones, even when the server is honest.

%
%
\section{Feature-space hijacking attack}
\label{sec:fsha}
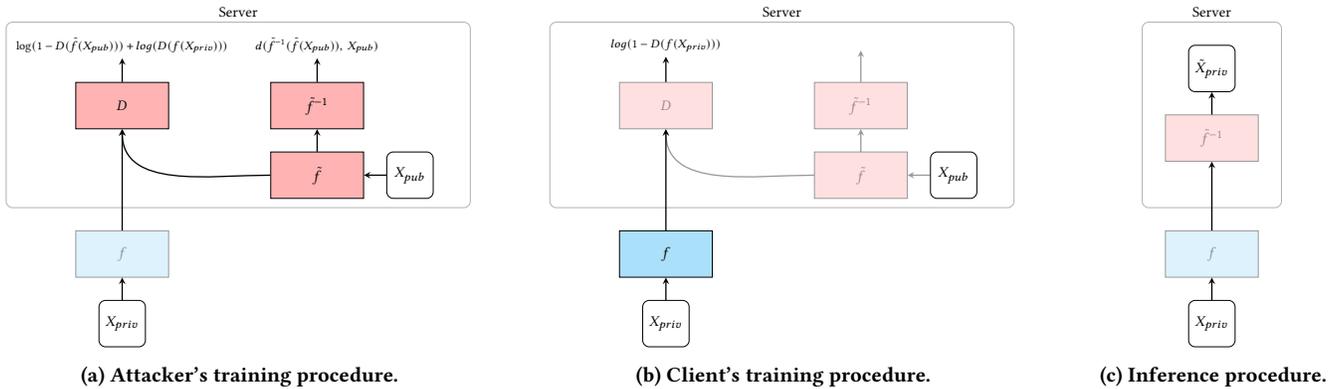
\begin{figure*}[t]
	\centering
	\begin{subfigure}{.35\textwidth}
		\centering
		\resizebox{1\textwidth}{!}{%
		\begin{tikzpicture}
		\tikzstyle{data} = [rectangle, rounded corners, minimum width=1cm, minimum height=1cm,text centered, draw=black, fill=white!30]
		\tikzstyle{net} = [rectangle, minimum width=2cm, minimum height=1cm,text centered, draw=black, fill=red!30]
		\tikzstyle{arrow} = [thick,->,>=stealth]
		\tikzstyle{gradient}  = [line width=1.5mm, green!40, opacity=.8,]
		
		%
		%
		%
		
		\node (sa)[rectangle, rounded corners, minimum width=10cm, minimum height=4cm,text centered, draw=black, fill=white, opacity=.3, yshift=4.5cm, xshift=-.5cm, label=above: Server] {};
		
		\node (0) [data, yshift=0, xshift=-3cm] {$\xpr$};
		\node (1) [net, above of=0, yshift=.5cm, fill=cyan!30, opacity=.4] {$\f$};
		
		\node (6) [net, above of=1, yshift=2.2cm] {$D$};
		
		\node (3) [data, above of=1, yshift=.7cm, xshift=6.2cm] {$\xpu$};
		\node (4) [net, left of=3, xshift=-1cm] {$\tf$};
		
		\node (2) [net, above of=4, yshift=.5cm] {$\tfi$};

		\node (5) [above of=2, yshift=.3cm] {\footnotesize$d(\tfi(\tf(\xpu)),\ \xpu)$};
		\node (7) [above of=6, yshift=.3cm] {\footnotesize$ \log(1-D(\tf(\xpu))) + log(D(\f(\xpr)))$};
		
		\node (inv)[rectangle, rounded corners, minimum width=1cm, minimum height=1cm,text centered, draw=black, fill=red!30, opacity=0, above of =1, xshift=-2cm, yshift=1cm] {};

		edge[bend right=90] node [left] {} (1);
		
		\draw [arrow] (0) -- (1);
		\draw [arrow] (1) -- (6);
		\draw [arrow] (3) -- (4);
		\draw [arrow] (2) -- (5);
		\draw [arrow] (4) -- (2);
		\draw [arrow] (6) -- (7);
		
		
		
		\draw  [->,black, thick, >=stealth] (4.west) to [out=180,in=-90] (6.south);

		
		\end{tikzpicture}
	}
	\caption{Attacker's training procedure.}
	\label{fig:attack_a}
	\end{subfigure}\hspace{1cm}\begin{subfigure}{.35\textwidth}
	\centering
\resizebox{1\textwidth}{!}{%
	\begin{tikzpicture}
	\tikzstyle{data} = [rectangle, rounded corners, minimum width=1cm, minimum height=1cm,text centered, draw=black, fill=white!30]
	\tikzstyle{net} = [rectangle, minimum width=2cm, minimum height=1cm,text centered, draw=black, fill=red!30]
	\tikzstyle{arrow} = [thick,->,>=stealth]
	\tikzstyle{gradient}  = [line width=1.5mm, green!40, opacity=.8,]
	
	%
	%
	%
	
	\node (sa)[rectangle, rounded corners, minimum width=10cm, minimum height=4cm,text centered, draw=black, fill=white, opacity=.3, yshift=4.5cm, xshift=-.5cm, label=above: Server] {};
	
	\node (0) [data, yshift=0, xshift=-3cm] {$\xpr$};
	\node (1) [net, above of=0, yshift=.5cm, fill=cyan!30] {$\f$};
	
	\node (6) [net, above of=1, yshift=2.2cm, opacity=.4] {$D$};
	
	\node (3) [data, above of=1, yshift=.7cm, xshift=6.2cm] {$\xpu$};
	\node (4) [net, left of=3, xshift=-1cm, opacity=.4] {$\tf$};
	
	\node (2) [net, above of=4, yshift=.5cm, opacity=.4] {$\tfi$};

	\node (5) [above of=2, yshift=.3cm, opacity=0] {};
	\node (7) [above of=6, yshift=.3cm] {\footnotesize$log(1-D(\f(\xpr)))$};
	
	\node (inv)[rectangle, rounded corners, minimum width=1cm, minimum height=1cm,text centered, draw=black, fill=red!30, opacity=0, above of =1, xshift=-2cm, yshift=1cm] {};

	edge[bend right=90] node [left] {} (1);
	
	\draw [arrow] (0) -- (1);
	\draw [arrow] (1) -- (6);
	\draw [arrow, opacity=.4] (3) -- (4);
	\draw [arrow, opacity=.4] (2) -- (5);
	\draw [arrow, opacity=.4] (4) -- (2);
	\draw [arrow] (6) -- (7);
	
	
	
	\draw  [->,black, thick, >=stealth, opacity=.4] (4.west) to [out=180,in=-90] (6.south);

	
	\end{tikzpicture}
}
	\caption{Client's training procedure.}
	\label{fig:attack_b}
\end{subfigure}\hspace{1cm}\begin{subfigure}{.18\textwidth}
\centering
\resizebox{.596\textwidth}{!}{%
\begin{tikzpicture}

\tikzstyle{data} = [rectangle, rounded corners, minimum width=1cm, minimum height=1cm,text centered, draw=black, fill=white!30]
\tikzstyle{net} = [rectangle, minimum width=2cm, minimum height=1cm,text centered, draw=black, fill=red!30]
\tikzstyle{arrow} = [thick,->,>=stealth]
\tikzstyle{gradient}  = [line width=1.5mm, green!40, opacity=.8,]

%
%
%

\node (sa)[rectangle, rounded corners, minimum width=3cm, minimum height=4cm,text centered, draw=black, fill=white, opacity=.3, yshift=4.5cm, xshift=0cm, label=above: Server] {};

\node (0) [data, yshift=0, xshift=0cm] {$\xpr$};
\node (1) [net, above of=0, yshift=.5cm, fill=cyan!30, opacity=.4] {$\f$};

\node (6) [net, above of=1, yshift=1.5cm, opacity=.4] {$\tfi$};
\node (7) [data, above of=6, yshift=.5cm] {$\txpr$};

\draw [arrow] (0) -- (1);
\draw [arrow] (1) -- (6);
\draw [arrow] (6) -- (7);

\end{tikzpicture}
}
	\caption{Inference procedure.}
\label{fig:attack_c}
\end{subfigure}

\caption{Schematic representation of the setup and inference process of the feature-space hijacking attack. In the scheme, opaque rectangles depict the neural networks actively taking part to the training. Instead, more transparent rectangles are networks that may participate to the forward propagation but do not modify their weights.}
\label{fig:attack}
\end{figure*}

Here, we introduce our main attack against the split learning training protocol---the Feature-space hijacking attack (FSHA). We start in Section~\ref{sec:th_model} by detailing the threat model. Then, Section~\ref{sec:attack_theory} introduces the core intuition behind the attack, as well as its formalization. Section~\ref{sec:attack_impl} covers the pragmatic aspects of the attack, demonstrating its effectiveness. Section~\ref{sec:att_inf} extends the FSHA framework to property inference attacks.
\subsection{Threat model}
\label{sec:th_model}
 We assume that the attacker does not have information on the clients participating in the distributed training, except those required to run the split learning protocol. The attacker has no information on the architecture of $\f$ and its weights. Moreover, the attacker ignores the task on which the distributed model is trained. However, the adversary knows a dataset $\xpu$ that captures the same domain of the clients' training sets (a \TT{shadow dataset}~\cite{mem}). For instance, if the model is trained on face images, $\xpu$ is also composed of face images. Nevertheless, no intersection between private training sets and $\xpu$ is required. This assumption is congruent with previous attacks against collaborative inference~\cite{model_inversion}, and makes our threat model more realistic and less restrictive than the ones adopted in other related works~\cite{splitnnsec, splitnnsecex}, where the adversary is assumed to have direct access to leaked pairs of \textit{smashed} data and private data.
\par

%
%
\subsection{Attack foundations}
\label{sec:attack_theory}
As discussed in Section~\ref{sec:splitnnsec}, the main vulnerability of split learning resides in the fact that the server has control over the learning process of the clients' network. Indeed, even ignoring the architecture of $\f$ and its weights, an adversary can forge a suitable gradient and force $\f$ to converge to an arbitrary target function chosen by the attacker. In doing so, the attacker can induce certain properties in the \textit{smashed data} generated by the clients, enabling inference or reconstruction attacks on the underlying private data.
\par

Here, we present a general framework that implements this attack procedure. In such a framework, the malicious server substitutes the original learning task chosen by the clients with a new objective that shapes, on purpose, the codomain/feature-space of~$\f$.\footnote{The client's network $\f$ can be seen as a mapping between a data space $\X$ (\ie where training instances are defined) and a feature space $\mathbf{Z}$ (\ie where smashed data are defined).} 
During the attack, the server exploits its control on the training process to hijack $\f$ and steer it towards a specific, target feature space $\Z$ that is appositely crafted. Once $\f$ maps into $\Z$, the attacker can recover the private training instances by locally inverting the known feature space.
\par

Such an attack encompasses two phases: (1)~a \textbf{setup phase} where the server hijacks the learning process of $\f$, and (2)~a subsequent \textbf{inference phase} where the server can freely recover the \textit{smashed} data sent from the clients. Hereafter, we refer to this procedure as \textbf{Feature-space Hijacking Attack}, FSHA for short.
\paragraph{Setup phase}
The setup phase occurs over multiple training iterations of split learning and is logically divided into two concurrent steps, which are depicted in Figures~\ref{fig:attack_a}~and~\ref{fig:attack_b}. In this phase of the attack, the server trains three different networks; namely, $\tf$, $\tfi$, and $D$. These serve very distinct roles; more precisely:
\begin{itemize}
	\item{$\tf$:} is a pilot network that dynamically defines the target feature space $\Z$ for the client's network $\f$. As $\f$, $\tf$ is a mapping between the data space and a target feature space $\Z$, where $|\tf(x)|=|\f(x)|=k$.
	\vspace{0.2cm}
	\item{$\tfi$:} is an approximation of the inverse function of $\tf$. During the training, we use it to guarantee the invertibility of~$\tf$ and recover the private data from \textit{smashed} data during the inference phase.
	\vspace{0.2cm}
	\item{$D$:} is a discriminator~\cite{gan} that indirectly guides $\f$ to learn a mapping between the private data and the feature space defined from $\tf$. Ultimately, this is the network that substitutes $\s$ in the protocol (\eg Figure~\ref{fig:split}), and that defines the gradient which is sent to the client during the distributed training process.
\end{itemize}
The setup procedure also requires an unlabeled dataset~$\xpu$ that is used to train the three attacker's networks. Observe that this is the only knowledge of the clients' setup that the attacker requires. The effect of $\xpu$ on the attack performance will be analyzed in the next section.
\par
As mentioned before, at every training iteration of split learning (\ie when a client sends \textit{smashed} data to the server), the malicious server trains the three networks in two concurrent steps, which are depicted in Figures~\ref{fig:attack_a}~and~\ref{fig:attack_b}. The server starts by sampling a batch from~$\xpu$ that uses to jointly train~$\tf$~and~$\tfi$. {Here, the server optimizes the weights of~$\tf$~and~$\tfi$ to make the networks converge towards an auto-encoding function \ie~$\tfi(\tf(x))=x$. This is achieved by minimizing the loss function:}
\begin{equation}
\loss_{\tf,\tfi} = d(\tfi(\tf(\xpu)),\ \xpu),
\label{eq:autoenc}
\end{equation}
where $d$ is a suitable distance function, \eg the Mean Squared Error~(MSE). 
\par

Concurrently, also the network $D$ is trained. \textbf{This is a discriminator~\cite{gan} that is trained to distinguish between the feature space induced from~$\tf$ and the one induced from the client's network~$\f$.} The network~$D$ takes as input $\tf(\xpu)$ or $\f(\xpr)$ (\ie the \textit{smashed} data) and is trained to assign high probability to the former and low probability to the latter. More formally, at each training iteration, the weights of~$D$ are tuned to minimize the following loss function:
\begin{equation}
\loss_{D} = \log(1-D(\tf(\xpu))) + \log(D(\f(\xpr))).
\end{equation}
After each local training step for $D$, the malicious server can then train the network $f$ by sending a suitable gradient signal to the remote client performing the training iteration. In particular, this gradient is forged by using $D$ to construct an adversarial loss function for $\f$; namely:
\begin{equation}
\loss_{\f}=\log(1-D(\f(\xpr))).
\end{equation}
That is, $\f$ is trained to maximize the probability of being miss-classified from the discriminator $D$. In other words, \textbf{we require the client's network to learn a mapping to a feature space that is indistinguishable from the one of $\tf$}. Ideally, this loss serves as a proxy for the more direct and optimal loss function:~$d(\f(x),\ \tf(x))$. However, the attacker has no control over the input of $\f$ and must overcome the lack of knowledge about $x$ by relying upon an adversarial training procedure that promotes a topological matching between feature spaces rather than a functional equivalence between networks.
\paragraph{Attack inference phase} After a suitable number of setup iterations, the network $\f$ reaches a state that allows the attacker to recover the private training instances from the \textit{smashed} data. Here, thanks to the adversarial training, the codomain of $\f$ \textbf{overlaps} with the one of $\tf$. The latter feature space is known to the attacker who can trivially recover $\xpr$ from the \textit{smashed} data by applying the inverse network $\tfi$. Indeed, as the network $\f$ is now mapping the data space into the feature space $\Z$, the network $\tfi$ can be used to map the feature space $\Z$ back to the data space, that is:
\[
\txpr = \tfi(\f(\xpr)),
\]
where $\txpr$ is a suitable approximation of the private training instances $\xpr$. This procedure is depicted in Figure~\ref{fig:attack_c}. The quality of the obtained reconstruction will be assessed later in the paper.
\par

We emphasize that the feature-space hijacking attack performs identically on the private-label version of the protocol, \eg Figure~\ref{fig:split_b}. In this case, at each training step, the server sends arbitrary forged inputs to the clients' final layers and ignores the gradient produced as a response, hijacking the learning process of $\f$ as in the previous instance. More generally, in the case of multiple vertical splits, a malicious party can always perform the attack despite its position in the stack. Basically, the attacker can just ignore the received gradient and replace it with the crafted one, leaving the underlying splits to propagate the injected adversarial task. Additionally, the effectiveness of the attack does not depend on the number of participating clients. 
\par 

In the same way, the feature-space hijacking attack equally applies to extensions of split learning such as \textit{Splitfed} learning~\cite{splitfed}. Indeed, in this protocol, the server still maintains control of the learning process of $f$. The only difference is in how the latter is updated and synchronized among the clients. Interestingly, the attack can be potentially more effective as the server receives bigger batches of \textit{smashed} data that can be used to smooth the learning process of the discriminator.
\par

In the next section, we implement the feature-space hijacking attack, and we empirically demonstrate its effectiveness on various setups.

\subsection{Attack implementations}
\label{sec:attack_impl}
\begin{figure*}
	\centering
		\resizebox{.75\textwidth}{!}{%
		\begin{tikzpicture}
		\tikzstyle{layer} = [rectangle, minimum width=2.8cm, minimum height=.6cm,text centered, draw=black, rotate=90]
		\tikzstyle{res} = [rectangle, minimum width=2.8cm, minimum height=.6cm,text centered, draw=black, rotate=90,  fill=red!30]
		
		\tikzstyle{level} = [text centered]		
		\tikzstyle{arrow} = [thick,->,>=stealth]
	
		\node (00)[ yshift=0cm, xshift=0cm,  fill=white] {$\xpr$};
	    \node (0) [layer,  yshift=-.5cm, below of=00, fill=blue!30] {\small \texttt{2D-Conv(64, 3,(1,1))}};
	    \node (1) [layer,  yshift=-.5cm, below of=0,  fill=red!30, minimum width=1.1cm] {\texttt{ReLU}};
	    \node (2) [layer,  yshift=-.5cm, below of=1,  fill=green!30] {\small \texttt{batch-normalization} };
	    \node (3) [layer,  yshift=-.5cm, below of=2,  fill=red!30, minimum width=1.1cm] {\texttt{ReLU}};
	    \node (4) [layer,  yshift=-.5cm, below of=3,  fill=blue!30] {\small \texttt{maxPolling((2,2))}};
	    \node (5) [res,  yshift=-.5cm, below of=4] {\small \texttt{resBlock(64,1)} };
	    
	     \node (l1) [level,  yshift=-2.2cm, left  of=5, xshift=1cm,] {\textbf{split 1}};
	    
	    \node (6) [res,  yshift=-.5cm, below of=5] {\small \texttt{resBlock(128,2)} };
	    \node (l2) [level,  yshift=-2.2cm, left  of=6, xshift=1cm,] {\textbf{split 2}};
	    
	    \node (7) [res,  yshift=-.5cm, below of=6] {\small \texttt{resBlock(128,1)} };
	    \node (l3) [level,  yshift=-2.2cm, left  of=7, xshift=1cm,] {\textbf{split 3}};
	    
	    \node (8) [res,  yshift=-.5cm, below of=7] {\small \texttt{resBlock(256,2)} };
	   \node (l4) [level,  yshift=-2.2cm, left  of=8, xshift=1cm,] {\textbf{split 4}};
	   
	     \node (9) [ xshift=.5cm, right of=8] {};

	    \draw [arrow] (00) -- (0);
	    \draw [arrow] (0) -- (1);
		\draw [arrow] (1) -- (2);

		\draw [arrow] (2) -- (3);
		\draw [arrow] (3) -- (4);

		\draw [arrow] (5) -- (l1);
		
		\draw [arrow] (4) -- (5);
		\draw [arrow] (5) -- (6);
		
		\draw [arrow] (6) -- (l2);
		
		\draw [arrow] (6) -- (7);
		\draw [arrow] (7) -- (l3);
			
		\draw [arrow] (7) -- (8);
		\draw [arrow] (8) -- (l4);
		\draw [dotted, thick,->,>=stealth] (8) -- (9);
		\end{tikzpicture}
	}
	\caption{Architecture of the client's network $\f$ divided in $4$ different depth levels. The internal setup of the adopted residual blocks is described in Algorithm~\ref{algo:rb} (Appendix~\ref{app:arch}).}
	\label{fig:farch}
\end{figure*}
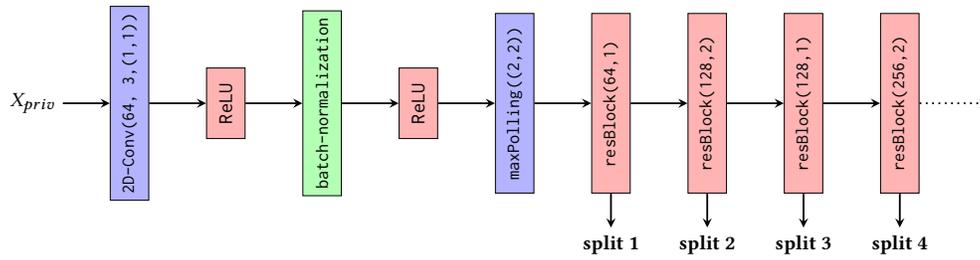
\begin{figure*}
	\centering
	\includegraphics[trim =13mm 75mm 0mm 0mm, clip, width=.4\linewidth]{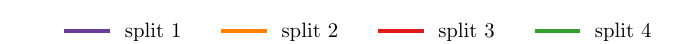}\\
	\begin{subfigure}{.24\textwidth}
		\centering
		\includegraphics[trim = 0mm 0mm 0mm 0mm, clip, width=1\linewidth]{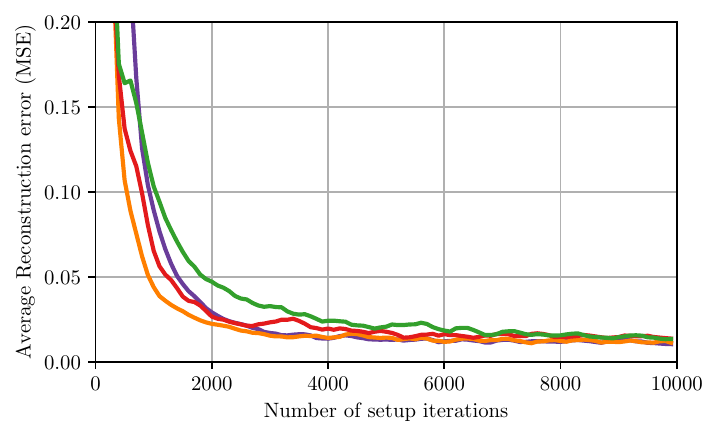}
		\caption{\textit{MNIST}.}\label{fig:levelsa}
	\end{subfigure} 
\begin{subfigure}{.24\textwidth}
		\centering
		\includegraphics[trim = 0mm 0mm 0mm 0mm, clip, width=1\linewidth]{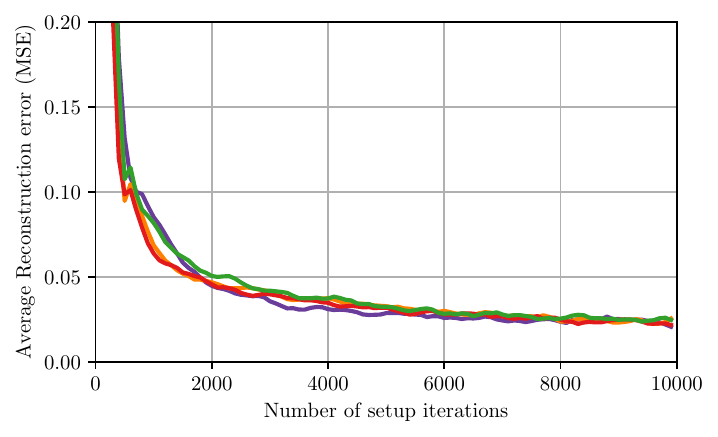}
		\caption{\textit{Fashion-MNIST}.}\label{fig:levelsb}
	\end{subfigure}
	\begin{subfigure}{.24\textwidth}
	\centering
	\includegraphics[trim = 0mm 0mm 0mm 0mm, clip, width=1\linewidth]{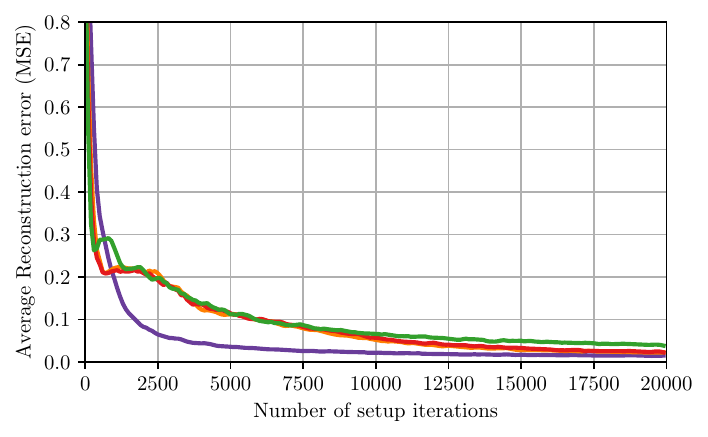}
	\caption{\textit{Omniglot}.}\label{fig:levelsc}
\end{subfigure} \begin{subfigure}{.24\textwidth}
	\centering
	\includegraphics[trim = 0mm 0mm 0mm 0mm, clip, width=1\linewidth]{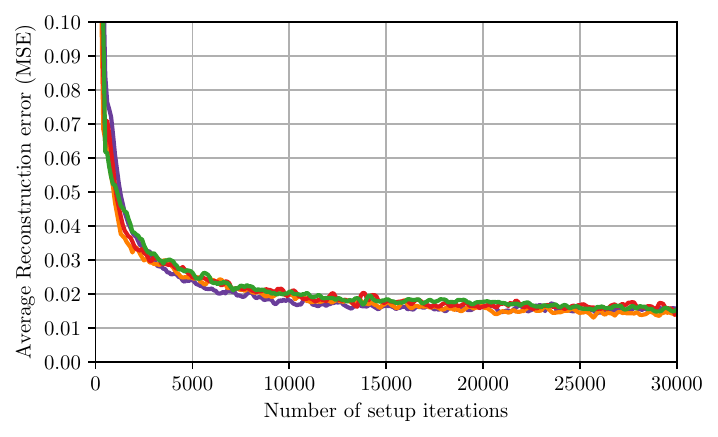}
	\caption{\textit{CelebA}.}\label{fig:levelsd}
\end{subfigure}

	\caption{Reconstruction error of private training instances during the setup phase for four different splits and four different datasets. This is measured as the average MSE between the images normalized in the $[-1,1]$ interval.}
	\label{fig:mse_rec}
\end{figure*}
\begin{figure*}
	\centering
	\fboxsep=0.01mm
	\fboxrule=0mm
	\begin{subfigure}{.95\linewidth}
		\centering
		\fcolorbox{black!50}{black!50}{\includegraphics[trim = 0mm 0mm 0mm 0mm, clip, width=1\linewidth]{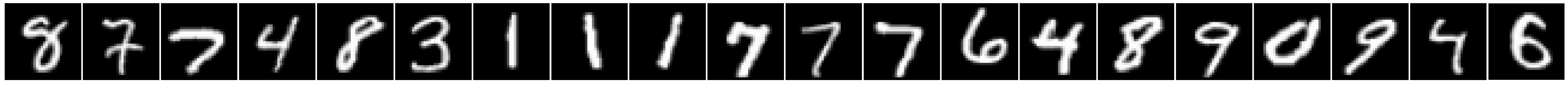}}\\
		\fcolorbox{red!50}{red!50}{\includegraphics[trim = 0mm 0mm 0mm 0mm, clip, width=1\linewidth, frame]{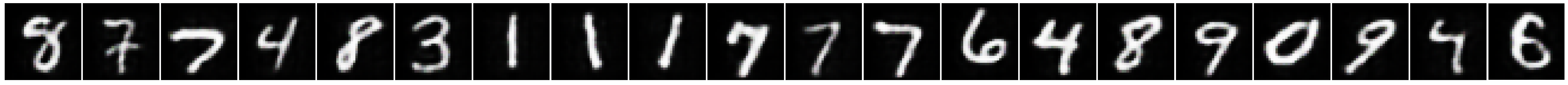}}
		\caption{MNIST.}
		\label{fig:rec_mnist}
	\end{subfigure}
	\begin{subfigure}{.95\linewidth}
		\centering
		\fcolorbox{black!50}{black!50}{\includegraphics[trim = 0mm 0mm 0mm 0mm, clip, width=1\linewidth]{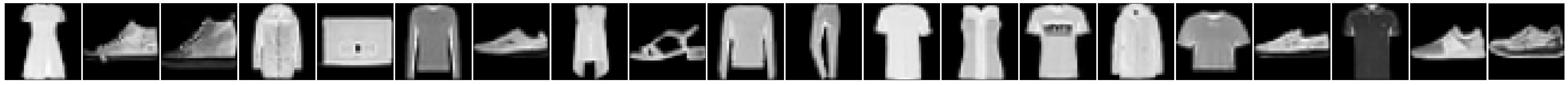}}\\
		\fcolorbox{red!50}{red!50}{\includegraphics[trim = 0mm 0mm 0mm 0mm, clip, width=1\linewidth]{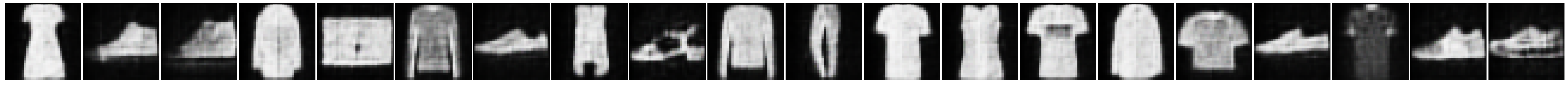}}
		\caption{Fashion-MNIST.}
	\label{fig:rec_fmnist}
\end{subfigure}

\begin{subfigure}{.95\linewidth}
	\centering
	\fcolorbox{black!50}{black!50}{\includegraphics[trim = 0mm 0mm 0mm 0mm, clip, width=1\linewidth]{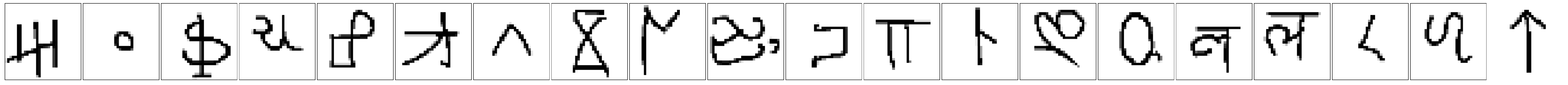}}\\
	\fcolorbox{red!50}{red!50}{\includegraphics[trim = 0mm 0mm 0mm 0mm, clip, width=1\linewidth]{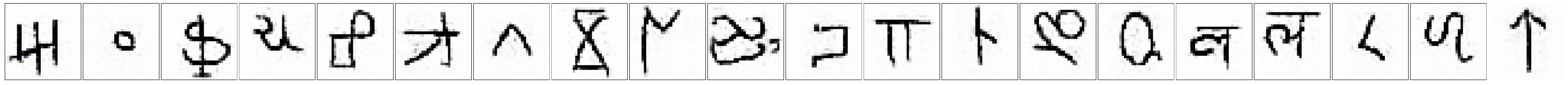}}
	\caption{Omniglot.}
	\label{fig:rec_omni}
\end{subfigure}
%
%
\begin{subfigure}{.95\linewidth}
	\centering
	\fcolorbox{black!50}{black!50}{\includegraphics[trim = 0mm 0mm 0mm 0mm, clip, width=1\linewidth]{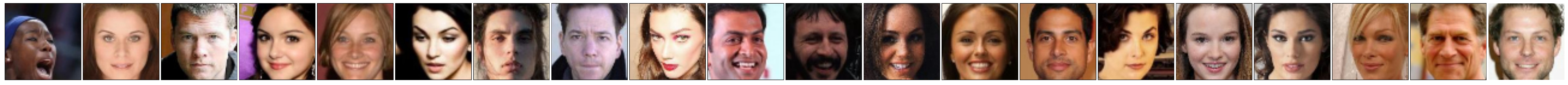}}\\
	\fcolorbox{red!50}{red!50}{\includegraphics[trim = 0mm 0mm 0mm 0mm, clip, width=1\linewidth]{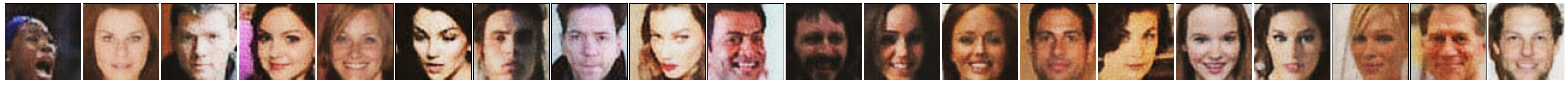}}
	\caption{CelebA.}
	\label{fig:rec_celeb}
\end{subfigure}
\caption{Examples of inference of private training instances from \textit{smashed data} for four datasets for the split $4$ of $\f$. Within each panel, the first row (\ie gray frame) reports the original data, whereas the second row (\ie red frame) depicts the attacker's reconstruction. The reported examples are chosen randomly from $\xpr$.}
\label{fig:rec_style}
\end{figure*}
\begin{figure}[b]
	\centering
	\includegraphics[trim = 0mm 0mm 0mm 0mm, clip, width=.95\linewidth]{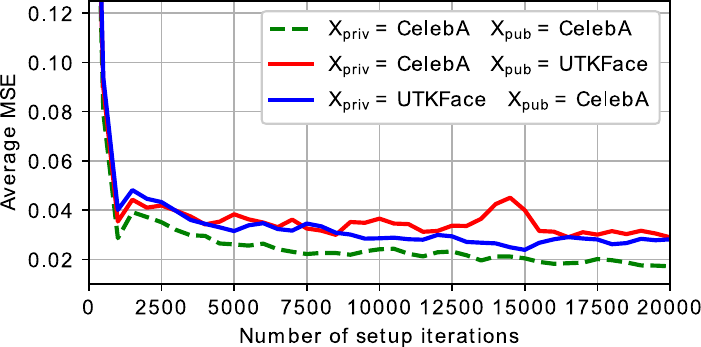}
	\caption{\new{Average reconstruction error during the FSHA for three different setups (zoomed in)}.}
	\label{fig:utkface_celeba_mse}
\end{figure}
\begin{figure*}
	\centering
	\fboxsep=0.01mm
	\fboxrule=0mm
	\centering
	
	\begin{subfigure}{.95\linewidth}
		\fcolorbox{black!50}{black!50}{\includegraphics[trim = 0mm 0mm 0mm 0mm, clip, width=1\linewidth]{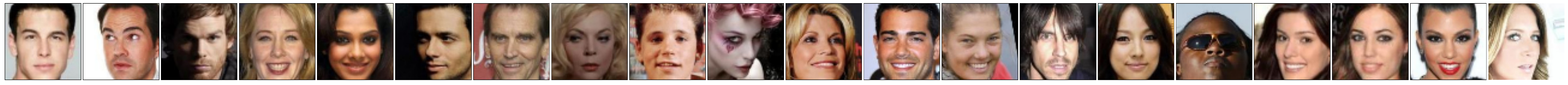}}\\
		\fcolorbox{red!50}{red!50}{\includegraphics[trim = 0mm 0mm 0mm 0mm, clip, width=1\linewidth, frame]{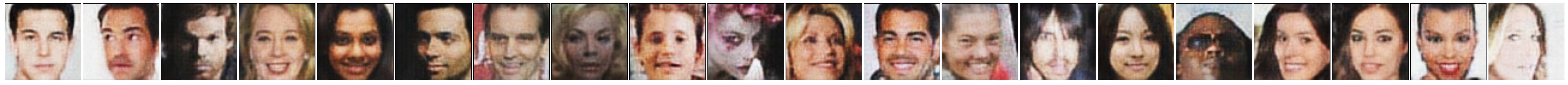}}
		\caption{Attacking \textit{CelebA} with $\xpu \myeq \textit{UTKFace}$.}
		\label{fig:utkface_celeba_a}
	\end{subfigure}\\
	
	\begin{subfigure}{.95\linewidth}
		\fcolorbox{black!50}{black!50}{\includegraphics[trim = 0mm 0mm 0mm 0mm, clip, width=1\linewidth]{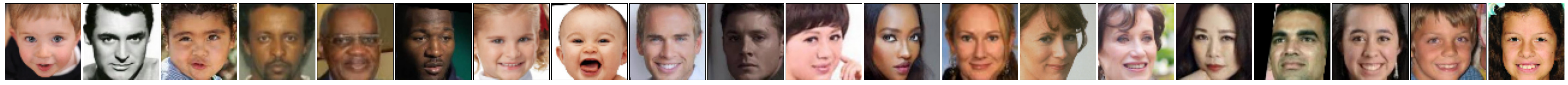}}\\
		\fcolorbox{red!50}{red!50}{\includegraphics[trim = 0mm 0mm 0mm 0mm, clip, width=1\linewidth, frame]{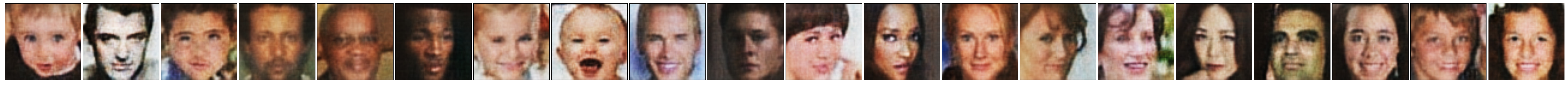}}
		\caption{Attacking \textit{UTKFace} with $\xpu \myeq \textit{CelebA}$.}
		\label{fig:utkface_celeba_b}
	\end{subfigure} 
	
	\caption{\new{Random examples of reconstruction attacks for two setups. Panel \textbf{(a)} reports the result for the case $\xpr \myeq \textit{CelebA}$ and $\xpu \myeq \textit{UTKFace}$. Panel \textbf{(b)} reports the result for the case $\xpr \myeq \textit{UTKFace}$ and $\xpu \myeq \textit{CelebA}$.}}
	\label{fig:utkface_celeba}
\end{figure*}

We focus on demonstrating the effectiveness of the attack on the image domain as this is predominant in split learning studies~\cite{splitnnsec, splitnnsecex, split_fed2, splitnn, splitnn_selfsurvey, splitfed, splitnn, splitnn2, splitnn3}. In our experiments, we rely on different image datasets to validate the attack; namely, \textit{MNIST}, Fashion-MNIST~\cite{fashionmnist}, \textit{Omniglot}~\cite{omni} and CelebA~\cite{celeba}. We demonstrate the effectiveness of the attack on additional datasets in Appendix~\ref{app:add_res_all}. In this section, we introduce the attack by simulating the clients' training set (\ie $\xpr$) using the training partition of the datasets, whereas we use their validation sets as $\xpu$. \new{Then, in Section~\ref{sec:public}, we demonstrate the effectiveness of the attack on subpar setups for the attacker.}
\paragraph{Attack setup}
We implement the various networks participating in the attack as deep convolution neural networks. For the client's network $\f$, we rely on a residual network~\cite{resnet} with a funnel structure---a pervasive architecture widely employed for tasks defined on the image domain. In our experiments, we test the proposed attack's effectiveness on increasingly deep splits of $\f$. These are depicted in Figure~\ref{fig:farch}.
\par

The network $\tf$ (the attacker's pilot network) is constructed \mb{by leveraging a different} architecture from the one used for $\f$. In particular, the network is chosen to be as simple as possible (\ie shallow and with a limited number of parameters). Intuitively, this allows us to define a very smooth target latent-space $\Z$ and simplify the learning process of $\f$ during the attack. The inverse mapping~$\tfi$ is \mb{also a shallow network} composed of transposed convolutional layers. The discriminator $D$ is a residual network and is chosen to be deeper than the other employed networks with the intent of forcing the feature spaces of $\f$ and $\tf$ to be as similar as possible until they become indistinguishable. During the setup phase, we regularize $D$ with a gradient penalty and use the Wasserstein loss~\cite{wgan} for the adversarial training. This greatly improves the stability of the attack and speeds up the convergence of $\f$. We rely on slightly different architectures for the attacker's networks (\ie $\tf$, $\tfi$~and~$D$) based on the depth of the split of $\f$. More detailed information about these, other hyper-parameters, and datasets pre-processing operations are given in Appendix~\ref{app:arch}. 
\par
%
%
%
%

\paragraph{Attack results}
During the attack, we use the MSE as the distance function~$d$~(see Eq.~\ref{eq:autoenc}). In the process, we track the attacker's reconstruction error measured as:
\[ MSE(\tfi(\f(\xpr)), \xpr).\] 
This is reported in Figure~\ref{fig:mse_rec} for the four datasets and four different splits of $\f$. In the experiments, different datasets required different numbers of setup iterations to reach adequate reconstructions. Low-entropy distributions \mb{like those in} \textit{MNIST} and \textit{Fashion-MNIST} can be accurately reconstructed within the first $10^{3}$ setup iterations. \new{On the other hand, natural images and complex distributions, like \textit{CelebA} and \textit{Omniglot}, tend to require more iterations ($3\cdot 10^{3}$ and $2\cdot 10^{3}$ respectively). It is important to note that clients depend entirely on the server and entrust it with the model's utility measure (validation error) during the training. The server, therefore, can directly control the stop-conditions (e.g., early-stopping) by providing suitable feedback to clients and dynamically requiring them to perform the number of training iterations needed to converge towards suitable reconstructions.}
\par

As the plots in Figure~\ref{fig:mse_rec} show, there is only a negligible difference in the reconstruction error achieved from attacks performed on the four different splits of~$\f$.
In those experiments, the depth of the client's network seems to affect only the convergence speed of the setup phase with a limited impact on the final performance.
\par

Also, in the case of the deepest split (split $4$), the FSHA allows the attacker to achieve precise reconstructions. These can be observed in Figure~\ref{fig:rec_style}, where the attack provides very accurate reconstructions of the original private data for all the tested datasets. More interestingly, the \textit{Omniglot} dataset highlights the generalization capability of the feature-space hijacking attack. The \textit{Omniglot} dataset is \mb{often} used as a benchmark for one-shot learning and contains $1623$ different classes with a limited number of examples each. The attack's performance on this dataset suggests that the proposed technique can reach a good generalization level over private data \mb{distributions}. We will investigate this property more thoroughly in the next section.\par

Hereafter, we will report results only for the split $4$ as this represents the worst-case scenario for our \mb{attack. Moreover,} it also captures the best practices of split learning. Indeed, deeper architectures for $\f$ are assumed to make it harder for an attacker to recover information from the \textit{smashed} data as this has been produced using more complex transformations~\cite{splitnn_what, splitnnsecex}.
\subsection{On the effect of the public dataset}
\label{sec:public}
It should be apparent that the training set $\xpu$ employed by the server can critically impact the attack's effectiveness. This is used to train the attacker's models and indirectly defines the target feature space imposed on $\f$. Ideally, to reach high-quality reconstruction, this should be distributed as similarly as possible to the private training sets owned by the clients. However, under strict assumptions, the attacker may collect data instances that are not sufficiently representative. We show that the Feature-space Hijacking Attack can be successfully applied even when the attacker employs inadequate/inaccurate choices of $\xpu$.
\subsubsection{Public dataset coming from a different distribution}
\label{sec:xpubdiff}
\new{Next, we analyze the effect of choices of $\xpu$ following a different distribution with respect to the private one. We start by attacking the dataset \textit{CelebA} ($\xpr$) relying on a different face dataset (\textit{UTKFace}~\cite{utkface}) as public dataset $\xpu$.}
\par

\new{
The \textit{UTKFace} dataset aggregates pictures of heterogeneous individuals. It portraits people within a wide age range (between $0$ to $116$ years) and covers several ethnicities (White, Black, Asian, Indian, and Others). The distribution of \textit{CelebA} is, instead, consistently more homogeneous and strongly skewed towards the Caucasian race with a stricter age range.\footnote{The dataset is composed of images of celebrities.}}
\par

\new{
We report the average reconstruction error of the attack together with the best-case scenario for the attacker (\textit{CelebA} train/validation partitions as $\xpr$ and $\xpu$) in Figure~\ref{fig:utkface_celeba_mse}. As can be observed, the discrepancy between private and public distributions affects the attack performance negligibly, and the FSHA can converge towards accurate reconstructions of private data. Figure~\ref{fig:utkface_celeba_a} depicts examples of such reconstructions.}
\par

\new{
For the sake of completeness, we report the results also for the opposite scenario: attacking \textit{UTKFace} with \textit{CelebA} as a public set. We obtained almost identical performance as shown in Figures~\ref{fig:utkface_celeba_mse} and \ref{fig:utkface_celeba_b}. Interestingly, in this case, the attack has also successfully reconstructed images of infants and gray-scale pictures that are missing in the \textit{CelebA} distribution.}
\par

\new{
In Appendix~\ref{app:add_res}, we repeat similar tests for the natural-image datasets \textit{TinyImageNet}~\cite{tiny} and \textit{STL-10}~\cite{stl10}, and dermatoscopic images datasets \textit{HAM10000}~\cite{ham} and \textit{ISIC-2016}~\cite{isic}, obtaining consistent results.}
\subsubsection{Public dataset with missing modalities}
\begin{figure}
	\includegraphics[trim = 0mm 0mm 0mm 0mm, clip, width=.9\linewidth]{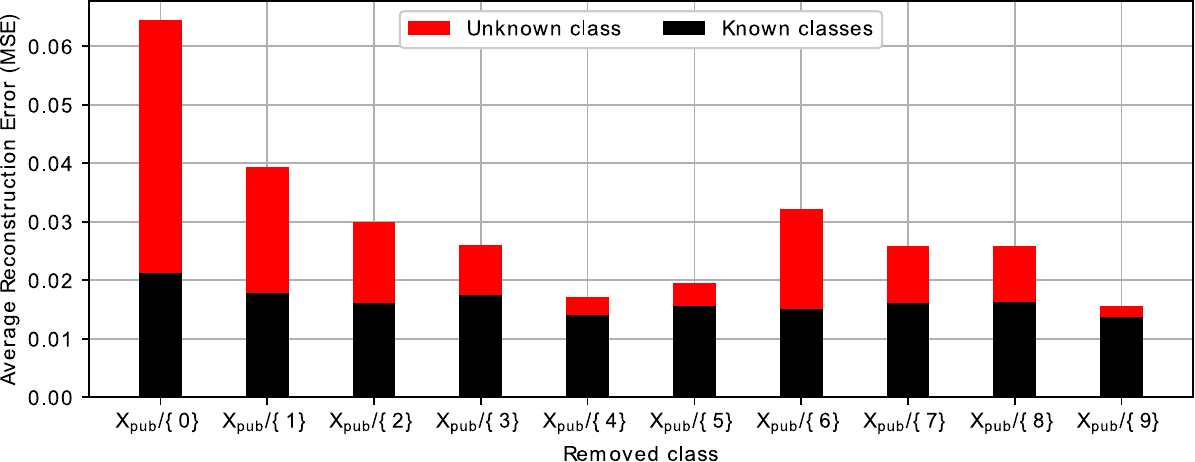}\\
	\caption{
		Each bar represents the final reconstruction error of private data obtained with an FSHA based on a $\xpu$ mangled of a specific class. \mb{Black bars report} the average reconstruction error of private data instances of classes known to the attacker. Instead, \mb{red bars report} the average reconstruction error of private data instances for the removed class. In the attacks, we used $15000$ setup iterations.}
	\label{fig:mangled_mse}
\end{figure}
\begin{figure}[b]
	\centering
	\fboxsep=0.01mm
	\fboxrule=0mm
	\centering
	\begin{subfigure}{.9\linewidth}
		\fcolorbox{black!50}{black!50}{\includegraphics[trim = 0mm 0mm 0mm 0mm, clip, width=1\linewidth]{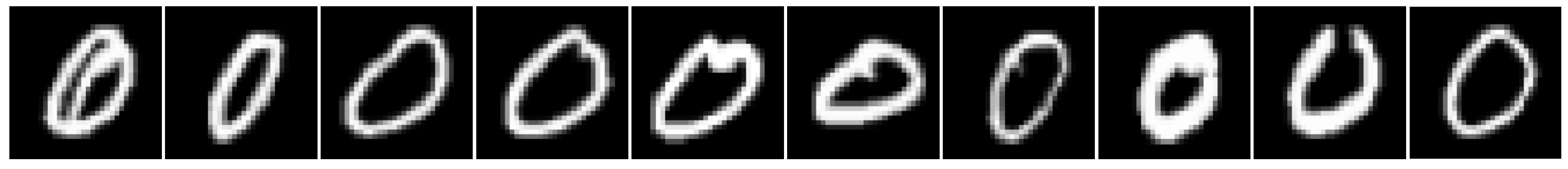}}\\
		\fcolorbox{red!50}{red!50}{\includegraphics[trim = 0mm 0mm 0mm 0mm, clip, width=1\linewidth, frame]{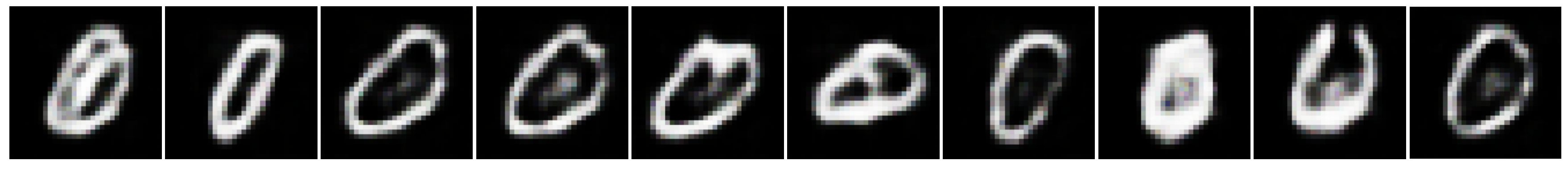}}
		\caption{Reconstruction $0$ with $\xpu/\{0\}$.}
	\end{subfigure}	\hspace{.1cm}\\
	\begin{subfigure}{.9\linewidth}
		\fcolorbox{black!50}{black!50}{\includegraphics[trim = 0mm 0mm 0mm 0mm, clip, width=1\linewidth]{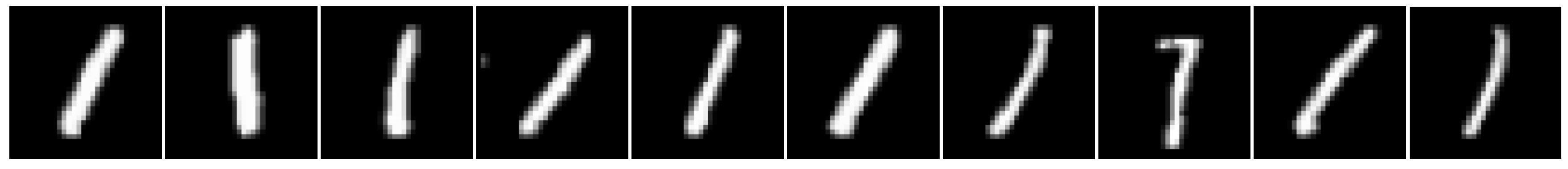}}\\
		\fcolorbox{red!50}{red!50}{\includegraphics[trim = 0mm 0mm 0mm 0mm, clip, width=1\linewidth, frame]{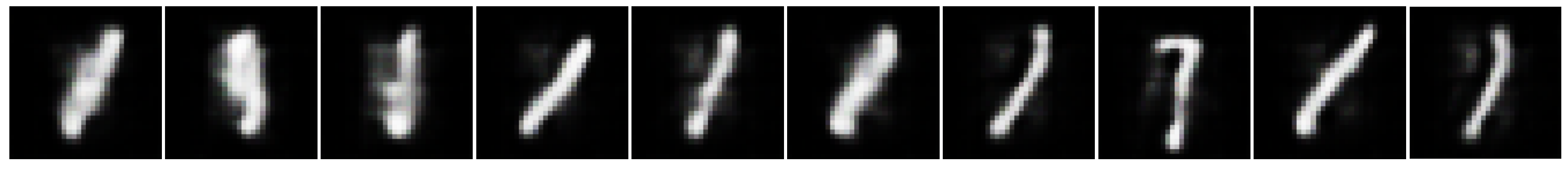}}
		\caption{Reconstruction $1$ with $\xpu/\{1\}$.}
	\end{subfigure}
	\caption{Two examples of inference of private training instances from \textit{smashed data} given mangled $\xpu$. In the panel~(a), the adversary carried out the attack without ever directly observing training instances representing the digit~\TT{$0$}. Panel~(b) reproduces the same result for the digit~\TT{$1$}. Only the reconstruction of instances of the class unknown to the attacker are reported. Those have been sampled from $\xpr$.}
	\label{fig:rec_mangled}
\end{figure}
Another interesting scenario is when the attacker's public set misses some modalities / semantic-classes of the private distribution. To simulate this scenario, we create artificially mangled training sets $\xpu$ for the \textit{MNIST} dataset and test the attack's effectiveness accordingly. In the experiment, the mangling operation removes all the instances of a specific class from $\xpu$ while leaving $\xpr$ (the training set used by the clients) unchanged. For instance, in the case of the \textit{MNIST} dataset, we remove from $\xpu$ all the images representing a specific digit. Then, we test the attack's ability to reconstruct instances of the removed class \ie data instances that the attacker has never observed during the setup phase. 
\par

Interestingly, the attack seems quite resilient to an incomplete $\xpu$. The results are \mb{depicted} in Figure~\ref{fig:rec_mangled} for $10$ different attacks carried out with $\xpu$ stripped of a specific class. For each attack, the average reconstitution error for the unknown classes (\ie red bars) is only slightly larger than the one for the classes represented in $\xpu$. Here, the attacker can successfully recover a suitable approximation of instances of the unobserved class by interpolating over the representations of observed instances. The only outlier is the case $\xpu/\{0\}$. Our explanation is that the digit zero is peculiar, so it is harder to describe it with a representation learned from the other digits. Nevertheless, as depicted in Figure~\ref{fig:rec_mangled}, the FSHA provides an accurate reconstruction also in the cases of $0$ and $1$.\\

\new{
Summing up, the public set leveraged by the attacker does impact the performance of the attack. Obviously, when the distribution of the public dataset is closer to the attacked one, it is possible to achieve a better reconstruction. However, as shown by the reported results, the attack is resilient to discrepancies of the public distribution, and it is capable of converging to precise reconstructions nonetheless. More interestingly, the attack procedure can generalize over unobserved modalities of the private distribution, allowing the attacker to leak suitable reconstructions of completely unobserved/unknown data classes. Eventually, these general properties of the attack make it applicable to realistic threat scenarios, where the adversary has just a limited knowledge of the target private sets.}

%
\subsection{Property inference attacks}
\label{sec:att_inf}
\begin{figure}
	\includegraphics[trim = 0mm 0mm 0mm 0mm, clip, width=1\linewidth]{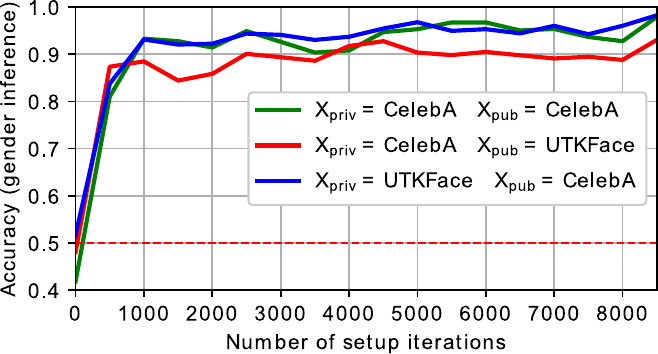}\\
	\caption{\new{Examples of property inference attack on the \textit{CelebA} and \textit{UTKFace} datasets. The plots report the accuracy in inferring the attribute \TT{gender} from instances of $\xpr$ during the setup phase of the attacks.}}
	\label{fig:acc_sex}
\end{figure}
%
%
%
In the previous setup, we demonstrated that it is possible to recover the entire input from the \textit{smashed} data. However, this type of inference may be sub-optimal for an attacker interested in inferring only a few specific attributes/properties of the private training instances (\eg the gender of the patients in medical records); rather than reconstructing $\xpr$ entirely. This form of inference was introduced in \cite{hackingsmartm} and extended to neural networks in \cite{ccs18_inference}. Property inference is simpler to perform and more robust to possible defensive mechanisms (see~Section~\ref{sec:def}).
Next, we briefly show how the Feature-space Hijacking Attack can be extended to perform property inference attacks.
\par

As discussed in Section~\ref{sec:attack_theory}, we can force arbitrary properties on the \textit{smashed} data produced by the clients by forging a tailored feature space $\Z$ and forcing the clients' network $\f$ to map into~it. The feature space $\Z$ is dynamically created by training a pilot network $\tf$ in a task that encodes the target property. In the attack of Figure~\ref{fig:attack}, we requested the invertibility of $\Z$ by training $\tf$ in an auto-encoding task with the support of a second network $\tfi$. Conversely, we can force the \textit{smashed} data to leak information about a specific attribute by conditioning the feature space $\Z$ with a classification task.
\par

It is enough to substitute the network $\tfi$ with a classifier $\fc$ trained to detect a particular attribute in the data points of $\Z$. However, unlike the previous formulation of the attack, the attacker has to resort to a supervised training set $(\xpu, \ypu)$ to define the target attribute. Namely, each instance of the attacker's dataset $\xpu$ must be associated with a label that expresses the attribute/property \textit{att} that the attacker wants to infer from the \textit{smashed} data. 
\par

In the case of a binary attribute, the attacker has to train $\fc$ in a binary classification using a binary cross-entropy loss function. 
Here, we implement the network $\fc$ to be as simple as possible to maximize the separability of the classes directly on $\Z$. In particular, we model $\fc$ as a linear model by using a single dense layer. In this way, we force the representations of the classes to be linearly separable, simplifying the inference attack once the adversarial loss has forced the topological equivalence between the codomains of $\f$ and $\tf$. We leave the other models and hyper-parameters unchanged.
\par 

In the experiments, we aim at inferring the binary attribute \textit{\TT{gender}} (\ie $0= $\TT{man}; $1= $\TT{woman}) from the private training instances used by the clients. \new{Following the results of Section~\ref{sec:xpubdiff}, we validate the proposed inference attack on different combinations of the datasets \textit{CelebA} and \textit{UTKFace} for $\xpr$ and $\xpu$.} During the attack, we track the accuracy of the inference attacks. \new{They are reported in Figure~\ref{fig:acc_sex}, where all the attacks reach an accuracy higher than $90\%$ within a limited number of iterations compared to the complete reconstruction attack. } 
\par

It is important to note that the property inference attack can be extended to any feature or task. For instance, the attacker can infer multiple attributes simultaneously by training $\fc$ in a multi-label classification rather than a binary one. The same applies to multi-class classification and regression tasks. In this direction, the only limitation is the attacker's capability to collect suitable labeled data to set up the attack. \new{Appendix~\ref{app:inf_cat} reports an additional example for a multi-class classification task.}
%
\subsection{Attack Implications}
The implemented attacks demonstrated how a malicious server could subvert the split learning protocol and infer information over the clients' private data. Here, the adversary can recover the single training instance from the clients and fully expose the distribution of the private data. Unlike previous attacks in collaborative learning~\cite{gan_attack, DLG}, the server can always determine exactly which client owns a training instance upon receiving the clients' disjointed \textit{smashed} data\footnote{\new{In split learning, the clients' activation cannot be aggregated.}}, further harming client's privacy.
\par

In the next section, we discuss the shortcomings of defense strategies proposed to prevent inference attacks.

\section{On defensive techniques}
\label{sec:def}
\begin{figure*}[t!]
	\centering
	\includegraphics[trim=0mm 90mm 150mm 0mm, clip, width=.8\linewidth]{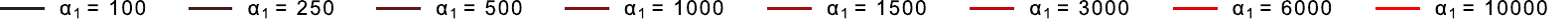}\\
	\begin{subfigure}{.5\textwidth}
		\centering
		\includegraphics[trim = 0mm 0mm 0mm 0mm, clip, width=.95\linewidth]{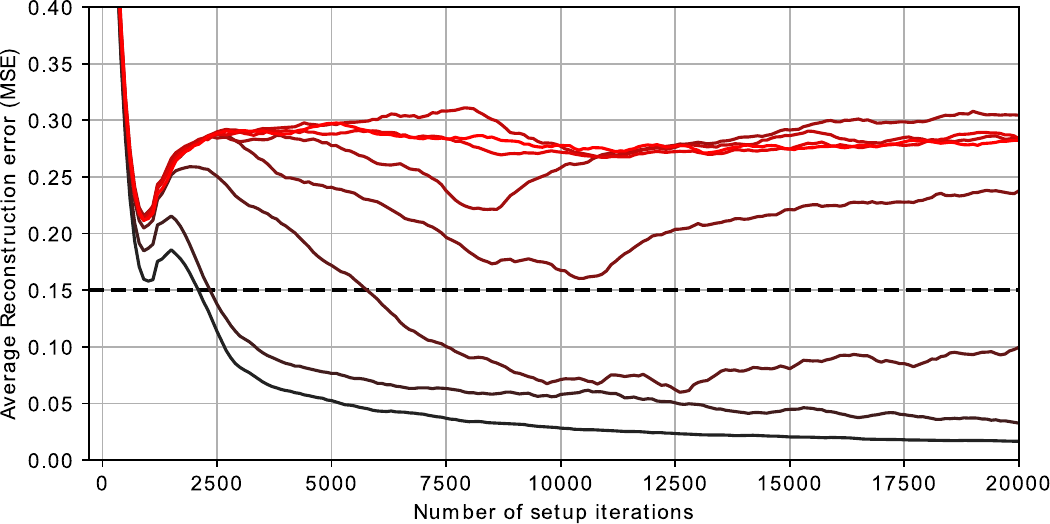}
		\caption{Default: task loss $\times1$}\label{fig:defa}
	\end{subfigure}\begin{subfigure}{.5\textwidth}
		\centering
		\includegraphics[trim = 0mm 0mm 0mm 0mm, clip, width=.95\linewidth]{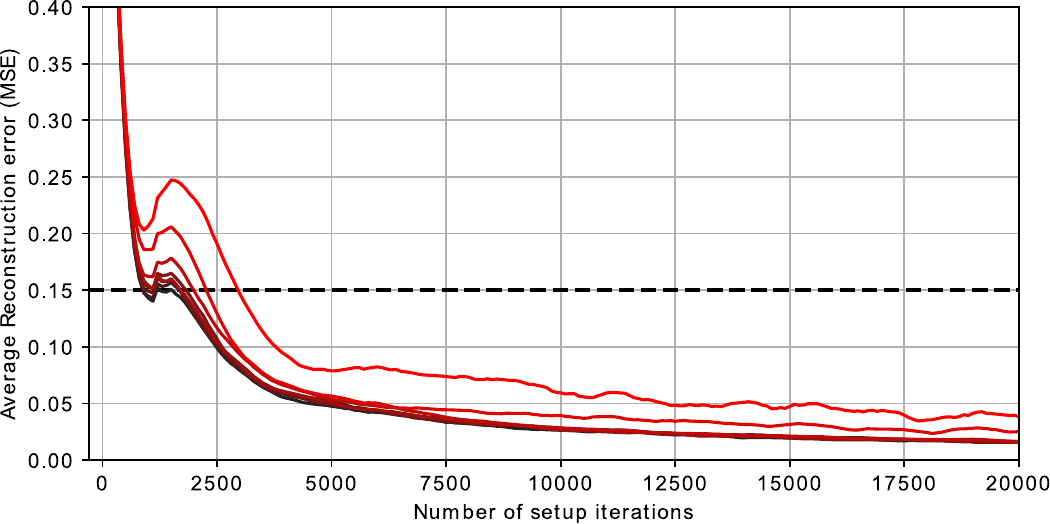}
		\caption{Re-weighted: task loss $\times25$.}\label{fig:defb}
	\end{subfigure}
	\caption{Effect of the distance correlation minimization defense on FSHA for the \textit{MNIST} dataset. Each curve in the figures depicts the reconstruction error of private data during the setup phase for a different value of $\alpha_1$ imposed by the client. The two panels report the effect of scaling the task loss (\eg $\alpha_2$) server-side.}
	\label{fig:def}
\end{figure*}
As demonstrated by our attacks, simply applying a set of neural layers over raw data cannot yield a suitable security level, especially when the adversary controls the learning process. As a matter of fact, as long as the attacker exerts influence on the target function of the clients' network, the latter can always be lead to insecure states. Unfortunately, there does not seem to be any way to prevent the server from controlling the learning process without rethinking the entire protocol from scratch. Next, we reason about the effectiveness of possible defense strategies.
\subsection{Distance correlation minimization}
In~\cite{splitnnsec, splitnnsecex}, the authors propose to artificially reduce the correlation between raw input and \textit{smashed} data by adding a regularization during the training of the distributed model in split learning. In particular, they resort to \textit{distance correlation}~\cite{distcorr}---a well-established measure of dependence between random vectors. Here, the clients optimize $\f$ to produce outputs that minimize the target task loss (\eg a classification loss) and the distance correlation. This regularization aims at preventing the propagation of information that is not necessary to the final learning task of the model from the private data to the \textit{smashed} one. Intuitively, this is supposed to hamper the reconstruction of $\xpr$ from an adversary that has access to the \textit{smashed} data.
\par

More formally, during the split learning protocol, the distributed model is trained to jointly minimize the following loss function: 
\begin{equation}
	\alpha_1 \cdot DCOR(\xpr, \f(\xpr))\ +\ \alpha_2 \cdot TASK(y, \s(\f(\xpr))),
	\label{eq:defloss}
\end{equation}
where $DCOR$ is the distance correlation metrics, $TASK$ is the task loss of the distributed model (\eg cross-entropy for a classification task), and $y$ is a suitable label for the target task (if any). In the equation, the hyper-parameters $\alpha_1$ and $\alpha_2$ define the relevance of distance correlation in the final loss function, creating and managing a tradeoff between data privacy (\ie how much information an attacker can recover from the smashed data) and model's utility on the target task (\eg the accuracy of the model in a classification task). Note that the distance correlation loss \mb{depends on just} the client's network $\f$ and the private data $\xpr$. Thus, it can be computed and applied locally on the client-side without any influence from the server.
\par

Even if the approach proposed in~\cite{splitnnsec, splitnnsecex} seems to offer reasonable security in the case of a passive adversary, it is, unfortunately, ineffective against the feature-space hijacking attack that influences the learning process of $\f$. \mb{As a matter of fact,} the learning objective injected by the attacker will naturally negate the distance correlation minimization, circumventing its effect. Moreover, this defensive technique does not prevent the property inference attack detailed in Section~\ref{sec:att_inf}.
\par

Figure~\ref{fig:defa} reports on the impact of the distance correlation minimization on the FSHA on the \textit{MNIST} dataset for different values of~$\alpha_1$. In the plot, we start from $\alpha_1~\myeq~100$, which is the smallest assignment of~$\alpha_1$ that does not affect the attack's performance, and we increase it until we reach impractical high values \eg $\alpha_1~\myeq~10000$. As shown in the \mb{plot}, the defense becomes effective when $\alpha_1$ reaches very high values. In these cases, the privacy loss completely eclipses the task loss of the distributed model (\ie Eq.~\ref{eq:defloss}). As a result, improving $\f$ in reducing the task loss becomes either impossible or extremely slow. Intuitively, this value of $\alpha_1$ prevents the distributed model from achieving any utility on the imposed task. This is so regardless of whether the model is trained on the task originally selected by the clients or the adversarial task enforced by the malicious server.
\par

Nevertheless, even if the clients set the parameter $\alpha_1$ to a large value, they have no effective method to control $\alpha_2$ if the server is malicious. 
Indeed, even in the label-private setting of split learning (\ie Figure~\ref{fig:split_b}), the server can arbitrarily determine the training objective for the model and adjust the task loss $TASK$. Trivially, this allows the attacker to indirectly control the ratio between the privacy loss (which is performed locally at the client) and the target loss (\ie the adversarial loss imposed by the attacker), nullifying the effect of a heavy regularization performed at the client-side. Figure~\ref{fig:defb} explicates how the malicious server circumvents the client-side defense by just scaling the adversarial loss function by a factor of $25$. In this case, even impractically large values of $\alpha_1$ are ineffective.
\par

To improve the defense mechanism above, one could apply gradient clipping on the gradient sent by the server during the training. However, gradient clipping further reduces the utility of the model as it weakens the contribution of the target loss function in the case of an honest server.
\par

Additionally, it is possible to devise a more general strategy and allow a malicious server to adopt advanced approaches to evade the defenses implemented in~\cite{splitnnsec, splitnnsecex}. Indeed, distance correlation can be easily circumvented by forging a suitable target feature space. The key idea is that the attacker can create an \TT{adversarial} feature space that minimizes the distance correlation \mb{ but allows the attacker to obtain a precise reconstruction of the input. We detail} this possibility in the Appendix~\ref{app:afsa}. Once the adversarial feature space is obtained, the attacker can hijack $\f$, minimize the distance correlation loss of $\f$, and recover the original data precisely. 

\subsection{Detecting the attack}
Alternatively, clients could detect the feature-space hijacking attack during the training phase and then halt the protocol. Unfortunately, detecting the setup phase of the attack seems to be a complex task due to the clients' incomplete knowledge of the distributed model. Here, clients could continuously test the effectiveness of the network on the original training task and figure out if the training objective has been hijacked. However, clients have no access to the full network during training and cannot query it to detect possible anomalies. This is also true for the private-label scenario, \ie Figure~\ref{fig:split_b} of split learning, where clients compute the loss function on their devices. Indeed, in this case, the attacker can simply provide fake inputs to $\f'$ (see Figure~\ref{fig:split_b}) that has been forged to minimize the clients' loss. For instance, the attacker can simply train a second dummy network $\tilde{\s}$ during the setup phase and send its output to the client. Here, the network $\tilde{\s}$ receives the \textit{smashed} data as input and is directly trained with the gradient received from $\f'$ to minimize the loss function chosen by the client. It's important to note that, during the attack, the network $\f$ does not receive the gradient from $\tilde{\s}$ but only from $D$.

\section{The security of split learning against malicious clients}
\label{sec:cattack}
\begin{algorithm}
	\KwData{Number of training iterations: $N$, Target class: $y_{t}$, Dummy class for poisoning $y_{\tilde{t}}$, Scaling factor gradient: $\epsilon$}
	\tcc{Initialize the local generative model}
	$G = \text{initGenerator}()$\;
	\For{$i$ \textbf{in} $[1, N]$}{
		\tcc{Download updated network splits}
		$\f, \f' = \text{get\_models()}$\;
		\tcc{Alterning poisoning attack and adversarial training}
		\tcc{(a more sophisticated scheduler may be used)}
		\If{$i\%2 == 0$}{
			$poisoning = True$\;
		}\Else{
			$poisoning = False$
		}
	
		\tcc{---- Start distributed forward-propagation}
		\tcc{Sample data instances from the generator $G$}
		$x \sim G$\;
		$z = \f(x)$\;
		\tcc{Send smashed data to the server and get $\s(\f(x))$ back}
		$z' = \text{send\_get\_forward(z)}$\;
		\tcc{Apply final layers and compute the probability for each class}
		$p = \f'(z')$\;
		\If{$poisoning$}{
			\tcc{Dummy label}
			$y= y_{\tilde{t}}$\;
		}\Else{
			\tcc{Target label}
			$y=y_{t}$\;
		}
		\tcc{Compute loss}
		$\loss = \text{cross-entropy}(y, p)$\;
		
		\tcc{---- Start distributed back-propagation}
		\tcc{Compute local gradient until $\s$}
		$\nabla_{\f^{'}} = \text{compute\_gradient}(\f^{'}, \loss)$\;
		\If{\textbf{not} $poisoning$}{
			\tcc{Scale down gradient}
			$\nabla_{\f^{'}} = \epsilon \cdot \nabla_{\f^{'}}$\;
		}\Else{
			\tcc{Apply gradient on $\f^{'}$}
			$\f^{'} = \text{apply}(\f^{'}, \nabla_{\f^{'}})$
		}
		\tcc{Send gradient to the server and receive gradient until $\f$}
		$\nabla_{\s} = \text{send\_get\_gradient}(\nabla_{\f^{'}})$\;
		\If{\textbf{not} $poisoning$}{
			\tcc{Scale back gradient}
			$\nabla_{\s} = \frac{1}{\epsilon} \cdot \nabla_{\s}$\;
		}
		\tcc{Compute local gradient until $G$}
		$\nabla_{\f} = \text{compute\_gradient}(\f, \nabla_{\s})$\;
		
		\If{$poisoning$}{
			\tcc{Apply gradient on $\f$}
			$\f = \text{apply}(\f, \nabla_{\f})$
		}\Else{
			\tcc{Compute local gradient until $G$'s input}
			$\nabla_{G} = \text{compute\_gradient}(G, \nabla_{\f})$\;
			\tcc{Apply gradient on the generator}
			$G = \text{apply}(G, \nabla_{G})$
		}
	}
	\caption{Client-side attack~\cite{gan_attack} in split learning.}
	\label{algo:gan_attack}
\end{algorithm}
In recent works~\cite{splitnn_selfsurvey}, the authors claim that the splitting methodology could prevent client-side attacks that were previously devised against federated learning, such as the GAN-based attack \cite{gan_attack}. Actually, we show that the attacks in \cite{gan_attack} (albeit with some minimal adaptations) remain applicable even within the split learning framework.
\paragraph{Client-side attack on Federated Learning}
The attack~\cite{gan_attack} works against the collaborative learning of a classifier $C$ trained to classify $n$ classes, say $y_1,\dots,\ y_n$. Here, a malicious client intends to reveal prototypical examples of a target class $y_{t}$, held by one or more honest clients. During the attack, the malicious client exerts control over a class $y_{\tilde{t}}$ that is used to actively poison the trained model and improve the reconstruction of instances $y_{t}$. 
\par

To perform the inference attack, the malicious client trains a local generative model~$G$ to generate instances of the target class~$y_{t}$. During each iteration, the attacker samples images from $G$, assigns the label~$y_{\tilde{t}}$ to these instances and uses them to train the model $C$ according to the learning protocol. Once the clients have contributed their training parameters, the attacker downloads the updated model~$C$ from the server and uses it as the discriminator~\cite{gan} to train the generative model $G$. The confidence of~$C$ on the class~$y_{t}$ is used as the discriminator's output and maximized in the loss function of~$G$. Once the generator has been trained, the attacker can use it to reproduce suitable target class instances $y_{t}$.
\begin{figure}[t]
	\centering
	\fboxsep=0.01mm
	\fboxrule=0mm
	\centering
	\begin{subfigure}{1\linewidth}
		\fcolorbox{red!50}{red!50}{\includegraphics[trim = 0mm 0mm 297mm 0mm, clip, width=1\linewidth, frame]{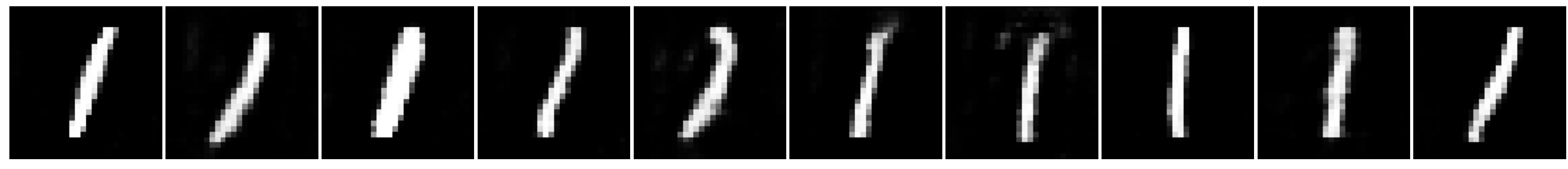}}
		\caption{\textit{MNIST} $y_{t} = 1$}
	\end{subfigure}\\
	\vspace{.1cm}
	
	\begin{subfigure}{1\linewidth}
	\fcolorbox{red!50}{red!50}{\includegraphics[trim = 0mm 0mm 297mm 0mm, clip, width=1\linewidth, frame]{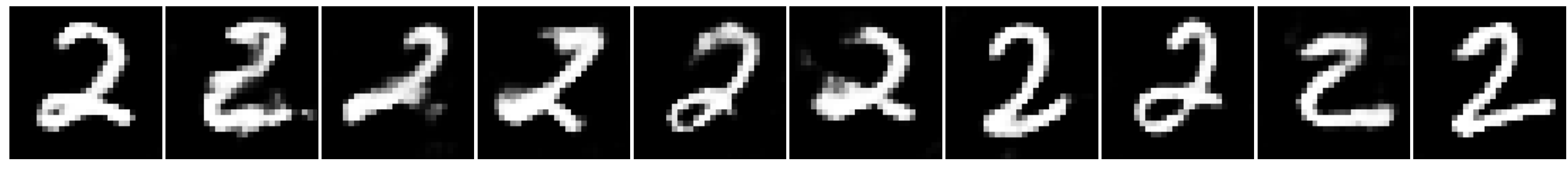}}
		\caption{\textit{MNIST} $y_{t} = 2$}
	\end{subfigure}
	\vspace{.1cm}
	
\begin{subfigure}{1\linewidth}
	\fcolorbox{red!50}{red!50}{\includegraphics[trim = 0mm 0mm 297mm 0mm, clip, width=1\linewidth, frame]{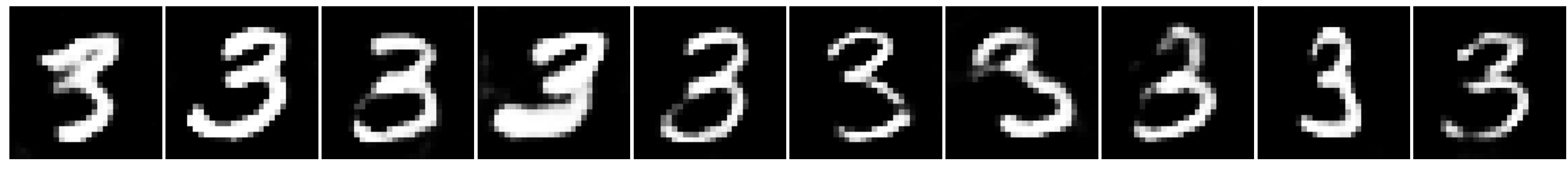}}
	\caption{\textit{MNIST} $y_{t} = 3$}
\end{subfigure}

\vspace{.6cm}

\begin{subfigure}{1\linewidth}
	\fcolorbox{black!50}{black!50}{\includegraphics[trim = 0mm 0mm 532mm 0mm, clip, width=.107\linewidth, frame]{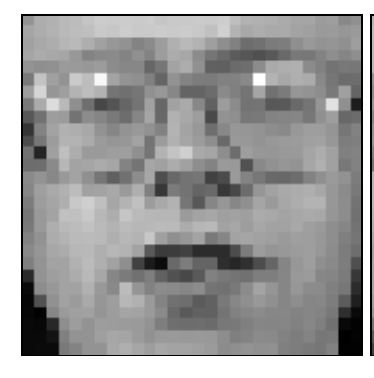}}
	\fcolorbox{red!50}{red!50}{\includegraphics[trim = 0mm 0mm 355mm 0mm, clip, width=.9\linewidth, frame]{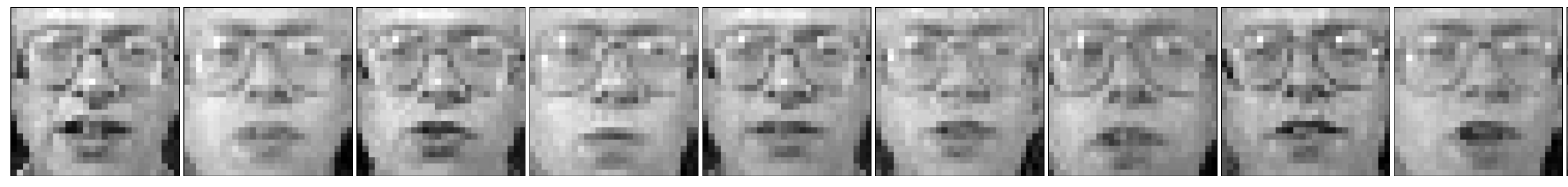}}
	\caption{\textit{AT\&T} $y_{t} = 1$}
\end{subfigure}\\
\vspace{.1cm}

\begin{subfigure}{1\linewidth}
	\fcolorbox{black!50}{black!50}{\includegraphics[trim = 0mm 0mm 532mm 0mm, clip, width=.107\linewidth, frame]{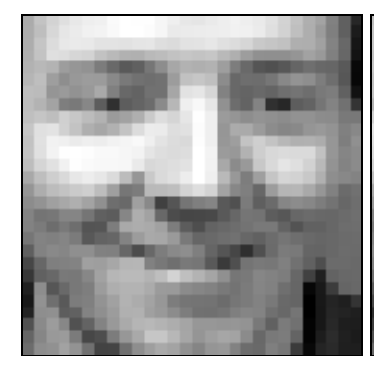}}
	\fcolorbox{red!50}{red!50}{\includegraphics[trim = 0mm 0mm 355mm 0mm, clip, width=.9\linewidth, frame]{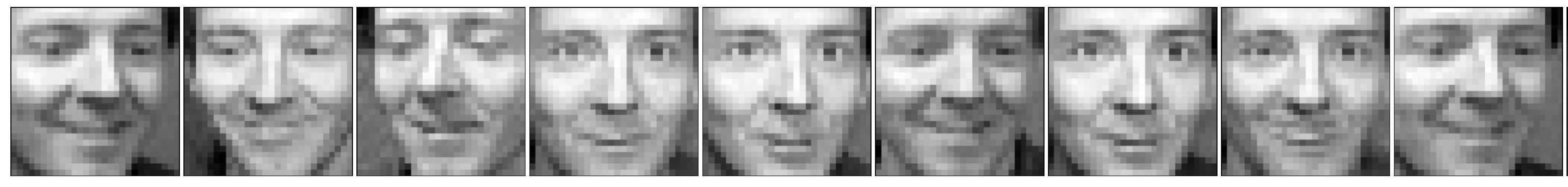}}
	\caption{\textit{AT\&T} $y_{t} = 2$}
\end{subfigure}
\vspace{.1cm}

\begin{subfigure}{1\linewidth}
	\fcolorbox{black!50}{black!50}{\includegraphics[trim = 0mm 0mm 532mm 0mm, clip, width=.107\linewidth, frame]{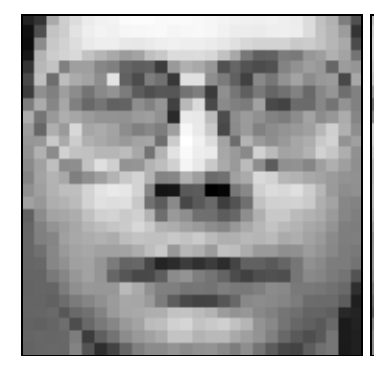}}
	\fcolorbox{red!50}{red!50}{\includegraphics[trim = 0mm 0mm 355mm 0mm, clip, width=.9\linewidth, frame]{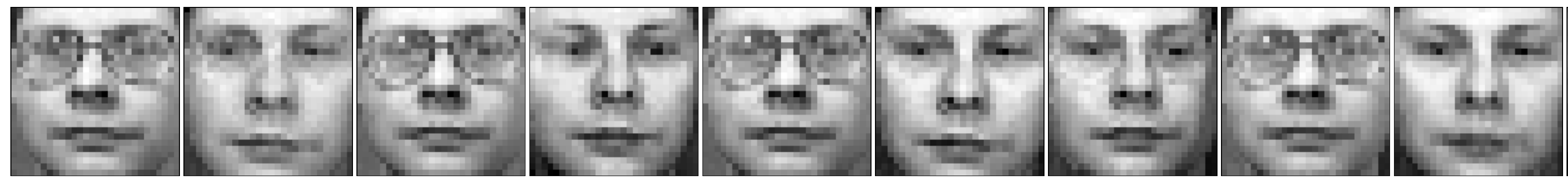}}
	\caption{\textit{AT\&T} $y_{t} = 3$}
\end{subfigure}
	\caption{Results from the client-side attack performed on split learning. The images are random samples from the generator trained via Algorithm~\ref{algo:gan_attack} on three attacks with different target classes. For the results on the dataset \textit{AT\&T}, we report also an instance of the target class in the leftmost corner of the panel in a gray frame.}
	\label{fig:client_gan_attack}
\end{figure}
\subsection{Client-side Attack on Split Learning}
The attack~\cite{gan_attack} can be performed on split learning under the same threat model. Note that, in this setup, the split learning server is honest, whereas the malicious client does not know the data distribution of the other clients' training sets.
\par

Considering the private-label case (\ie Figure~\ref{fig:split_b}), a malicious client exerts a strong influence over the learning process of the shared model $C=\f^{'}(\s(\f(\cdot))$ and can set up an attack similar to the one performed on federated learning. Here, the attacker trains a generator $G$ by using the distributed model $C=\f^{'}(\s(\f(\cdot)))$ as the discriminator by just providing suitable pairs (input, label) during the split learning protocol. This attack procedure is summarized in Algorithm~\ref{algo:gan_attack}. During the attack, the only impediment is the limited control of the attacker on the weights update procedure of the network $s$ hosted by the server. \textbf{Indeed, to soundly train the generator using the adversarial loss based on the distributed model $C$, the attacker must prevent the update of $s$ while training the generator~$G$}. However, the weights update operation of $s$ is performed by the server and cannot be directly prevented by the malicious client.\footnote{In this case, the back-propagation is performed client-side, and the malicious client can explicitly avoid updating the weights.}
\paragraph{The gradient-scaling trick} Nevertheless, this limitation can be easily circumvented by manipulating the gradient sent and received by the server during the split learning protocol. In particular, the malicious client can resort to gradient-scaling to make the training operation's impact on $s$ negligible. Here, before sending the gradient~$\nabla_{\f^{'}}$ produced from~$\f^{'}$ to~$s$, the client can multiply $\nabla_{\f^{'}}$ by a very small constant~$\epsilon$; that is:
\begin{equation}
 \nabla_{\f^{'}} = \epsilon \cdot \nabla_{\f^{'}}.
\end{equation}
This operation makes the magnitude of $\nabla_{\f^{'}}$, and so the magnitude of the weights update derived from it on $s$, negligible, \textbf{thus preventing any functional change in the weights of $s$}. Ideally, this is equivalent to force the server to train $s$ with a learning rate close to zero.
\par

Then, once $s$ has performed its back-propagation step and sent the gradient $\nabla_s$ to $\f$, the malicious client scales back $\nabla_s$ to its original magnitude by multiplying it by the inverse of $\epsilon$; that is:

\begin{equation}
\nabla_{\s} = \frac{1}{\epsilon} \cdot \nabla_{\s}.
\end{equation}

This allows the attacker to recover a suitable training signal for the generator $G$ that follows the back-propagation chain. Note that the malicious client does not update the weights of $\f$ or those of $\f'$ in the process. Eventually, the gradient-scaling operation allows the malicious client to train the generator using the distribute model $C$ as a discriminator. We demonstrate the soundness of this procedure later in this section.
\par

Although the gradient-scaling trick may provide a cognizant server an easy way to detect the attackers, a malicious client can always find a trade-off between attack secrecy and attack performance by choosing suitable assignments of $\epsilon$. As a matter of fact, it is hard for the server to distinguish the scaled gradient from the one achieved by a batch of {\em easy examples} (that is, data instances that the model correctly classifies with high confidence.)\\

The poisoning step of the attack~\cite{gan_attack} can be performed without any modification. The malicious client has to assign the label $y_{\tilde{t}}$ to instances sampled from the generator $G$ and run the standard split learning training procedure. In this process, the attacker updates the weights of all the participating networks but $G$. However, during the attack, the malicious client must alternate between a poisoning step and a \mb{genuine} training iteration for the generator as these cannot be performed simultaneously due to the gradient-scaling trick required to train the generator. Alternatively, the attacker can impersonate an additional client in the protocol and perform the poisoning iterations separately.
	
%
%
\paragraph{Attack validation}
	To implement the attack, we rely on architectures and hyper-parameters \mb{compatible with those originally used in~\cite{gan_attack} and perform the attack on the \textit{MNIST} and \textit{AT\&T} datasets.} More details are given in Appendix \ref{app:csa}. We use $\epsilon \myeq 10^{-5}$ in the \TT{gradient-scaling trick}. 
	In our setup, we model $10$ honest clients and a single malicious client who performs the attack described in Algorithm~\ref{algo:gan_attack}. In the process, we use the standard sequential training procedure of split learning~\cite{splitnn}. However, the attack equally applies to parallel extensions such as \textit{Splitfed} learning~\cite{splitfed}. We run the attack for $10000$ global training iterations. The results are reported in Figure~\ref{fig:client_gan_attack} for three attacks targeting different $y_{t}$, and prove the generator is successfully reproducing instances of the target class.

\section{Final Remarks}
\label{sec:conc}

In the present work, we described various structural vulnerabilities of split learning and showed how to exploit them and violate the protocol's privacy-preserving property. Here, an attacker can accurately reconstruct, or infer properties on, training instances. Additionally, we have shown that defensive techniques devised to protect split learning can be easily evaded.
\par

While federated learning exhibits similar vulnerabilities, split learning appears worse since it consistently leaks more information. Furthermore, it makes it even harder to detect ongoing inference attacks. 
Indeed, in standard federated learning, all participants store the neural network in its entirety, enabling simple detection mechanisms that, if nothing else, can thwart unsophisticated attacks.
%
%
%
%
\section*{Acknowledgments}
We acknowledge the generous support of \textit{Accenture} and the collaboration with their Labs in \textit{Sophia Antipolis}.
\bibliographystyle{ACM-Reference-Format}
\bibliography{bib}


\begin{thebibliography}{00}


\ifx \showCODEN    \undefined \def \showCODEN     #1{\unskip}     \fi
\ifx \showDOI      \undefined \def \showDOI       #1{#1}\fi
\ifx \showISBNx    \undefined \def \showISBNx     #1{\unskip}     \fi
\ifx \showISBNxiii \undefined \def \showISBNxiii  #1{\unskip}     \fi
\ifx \showISSN     \undefined \def \showISSN      #1{\unskip}     \fi
\ifx \showLCCN     \undefined \def \showLCCN      #1{\unskip}     \fi
\ifx \shownote     \undefined \def \shownote      #1{#1}          \fi
\ifx \showarticletitle \undefined \def \showarticletitle #1{#1}   \fi
\ifx \showURL      \undefined \def \showURL       {\relax}        \fi
\providecommand\bibfield[2]{#2}
\providecommand\bibinfo[2]{#2}
\providecommand\natexlab[1]{#1}
\providecommand\showeprint[2][]{arXiv:#2}

\bibitem[\protect\citeauthoryear{??}{spl}{2020}]%
        {split_com0}
 \bibinfo{year}{2020}\natexlab{}.
\newblock \bibinfo{title}{OpenMined: SplitNN}.
\newblock
  \bibinfo{howpublished}{\url{https://blog.openmined.org/tag/splitnn/}}.
  (\bibinfo{year}{2020}).
\newblock


\bibitem[\protect\citeauthoryear{??}{SLD}{2021}]%
        {SLDML}
 \bibinfo{year}{2021}\natexlab{}.
\newblock \bibinfo{title}{Workshop on Split Learning for Distributed Machine
  Learning (SLDML’21)}.
\newblock
  \bibinfo{howpublished}{\url{https://splitlearning.github.io/workshop.html}}.
   (\bibinfo{year}{2021}).
\newblock


\bibitem[\protect\citeauthoryear{Abadi, Chu, Goodfellow, McMahan, Mironov,
  Talwar, and Zhang}{Abadi et~al\mbox{.}}{2016}]%
        {dp}
\bibfield{author}{\bibinfo{person}{Martin Abadi}, \bibinfo{person}{Andy Chu},
  \bibinfo{person}{Ian Goodfellow}, \bibinfo{person}{H.~Brendan McMahan},
  \bibinfo{person}{Ilya Mironov}, \bibinfo{person}{Kunal Talwar}, {and}
  \bibinfo{person}{Li Zhang}.} \bibinfo{year}{2016}\natexlab{}.
\newblock \showarticletitle{Deep Learning with Differential Privacy}. In
  \bibinfo{booktitle}{{\em Proceedings of the 2016 ACM SIGSAC Conference on
  Computer and Communications Security}} {\em (\bibinfo{series}{CCS '16})}.
  \bibinfo{publisher}{Association for Computing Machinery},
  \bibinfo{address}{New York, NY, USA}, \bibinfo{pages}{308–318}.
\newblock
\showISBNx{9781450341394}
\showDOI{%
\url{https://doi.org/10.1145/2976749.2978318}}


\bibitem[\protect\citeauthoryear{Abedi and Khan}{Abedi and Khan}{2020}]%
        {splitrnn}
\bibfield{author}{\bibinfo{person}{Ali Abedi} {and} \bibinfo{person}{Shehroz~S.
  Khan}.} \bibinfo{year}{2020}\natexlab{}.
\newblock \bibinfo{title}{FedSL: Federated Split Learning on Distributed
  Sequential Data in Recurrent Neural Networks}.
\newblock   (\bibinfo{year}{2020}).
\newblock
\showeprint[arxiv]{cs.LG/2011.03180}


\bibitem[\protect\citeauthoryear{Abuadbba, Kim, Kim, Thapa, Camtepe, Gao, Kim,
  and Nepal}{Abuadbba et~al\mbox{.}}{2020}]%
        {splitnn_what}
\bibfield{author}{\bibinfo{person}{Sharif Abuadbba}, \bibinfo{person}{Kyuyeon
  Kim}, \bibinfo{person}{Minki Kim}, \bibinfo{person}{Chandra Thapa},
  \bibinfo{person}{Seyit~A. Camtepe}, \bibinfo{person}{Yansong Gao},
  \bibinfo{person}{Hyoungshick Kim}, {and} \bibinfo{person}{Surya Nepal}.}
  \bibinfo{year}{2020}\natexlab{}.
\newblock \bibinfo{title}{Can We Use Split Learning on 1D CNN Models for
  Privacy Preserving Training?}
\newblock   (\bibinfo{year}{2020}).
\newblock
\showeprint[arxiv]{cs.CR/2003.12365}


\bibitem[\protect\citeauthoryear{{Adam James Hall}}{{Adam James Hall}}{2020}]%
        {split_com1}
\bibfield{author}{\bibinfo{person}{{Adam James Hall}}.}
  \bibinfo{year}{2020}\natexlab{}.
\newblock \bibinfo{title}{Split Neural Networks on PySyft}.
\newblock
  \bibinfo{howpublished}{\url{https://medium.com/analytics-vidhya/split-neural-networks-on-pysyft-ed2abf6385c0}}.
    (\bibinfo{year}{2020}).
\newblock


\bibitem[\protect\citeauthoryear{Annas}{Annas}{2003}]%
        {PMID:12686707}
\bibfield{author}{\bibinfo{person}{George~J Annas}.}
  \bibinfo{year}{2003}\natexlab{}.
\newblock \showarticletitle{HIPAA regulations - a new era of medical-record
  privacy?}
\newblock \bibinfo{journal}{{\em The New England journal of medicine\/}}
  \bibinfo{volume}{348}, \bibinfo{number}{15} (\bibinfo{date}{April}
  \bibinfo{year}{2003}), \bibinfo{pages}{1486—1490}.
\newblock
\showISSN{0028-4793}
\showDOI{%
\url{https://doi.org/10.1056/nejmlim035027}}


\bibitem[\protect\citeauthoryear{Ateniese, Mancini, Spognardi, Villani, Vitali,
  and Felici}{Ateniese et~al\mbox{.}}{2015}]%
        {hackingsmartm}
\bibfield{author}{\bibinfo{person}{Giuseppe Ateniese},
  \bibinfo{person}{Luigi~V. Mancini}, \bibinfo{person}{Angelo Spognardi},
  \bibinfo{person}{Antonio Villani}, \bibinfo{person}{Domenico Vitali}, {and}
  \bibinfo{person}{Giovanni Felici}.} \bibinfo{year}{2015}\natexlab{}.
\newblock \showarticletitle{Hacking Smart Machines with Smarter Ones: How to
  Extract Meaningful Data from Machine Learning Classifiers}.
\newblock \bibinfo{journal}{{\em Int. J. Secur. Netw.\/}} \bibinfo{volume}{10},
  \bibinfo{number}{3} (\bibinfo{date}{Sept.} \bibinfo{year}{2015}),
  \bibinfo{pages}{137–150}.
\newblock
\showISSN{1747-8405}
\showDOI{%
\url{https://doi.org/10.1504/IJSN.2015.071829}}


\bibitem[\protect\citeauthoryear{Bagdasaryan, Veit, Hua, Estrin, and
  Shmatikov}{Bagdasaryan et~al\mbox{.}}{2020}]%
        {fedbackdoor}
\bibfield{author}{\bibinfo{person}{Eugene Bagdasaryan},
  \bibinfo{person}{Andreas Veit}, \bibinfo{person}{Yiqing Hua},
  \bibinfo{person}{Deborah Estrin}, {and} \bibinfo{person}{Vitaly Shmatikov}.}
  \bibinfo{year}{2020}\natexlab{}.
\newblock \showarticletitle{How To Backdoor Federated Learning}. In
  \bibinfo{booktitle}{{\em Proceedings of the Twenty Third International
  Conference on Artificial Intelligence and Statistics}} {\em
  (\bibinfo{series}{Proceedings of Machine Learning Research})},
  \bibfield{editor}{\bibinfo{person}{Silvia Chiappa} {and}
  \bibinfo{person}{Roberto Calandra}} (Eds.), Vol.~\bibinfo{volume}{108}.
  \bibinfo{publisher}{PMLR}, \bibinfo{address}{Online},
  \bibinfo{pages}{2938--2948}.
\newblock
\showURL{%
\url{http://proceedings.mlr.press/v108/bagdasaryan20a.html}}


\bibitem[\protect\citeauthoryear{Bhagoji, Chakraborty, Mittal, and
  Calo}{Bhagoji et~al\mbox{.}}{2019}]%
        {fed_poison0}
\bibfield{author}{\bibinfo{person}{Arjun~Nitin Bhagoji},
  \bibinfo{person}{Supriyo Chakraborty}, \bibinfo{person}{Prateek Mittal},
  {and} \bibinfo{person}{Seraphin Calo}.} \bibinfo{year}{2019}\natexlab{}.
\newblock \showarticletitle{Analyzing Federated Learning through an Adversarial
  Lens} {\em (\bibinfo{series}{Proceedings of Machine Learning Research})},
  \bibfield{editor}{\bibinfo{person}{Kamalika Chaudhuri} {and}
  \bibinfo{person}{Ruslan Salakhutdinov}} (Eds.), Vol.~\bibinfo{volume}{97}.
  \bibinfo{publisher}{PMLR}, \bibinfo{address}{Long Beach, California, USA},
  \bibinfo{pages}{634--643}.
\newblock
\showURL{%
\url{http://proceedings.mlr.press/v97/bhagoji19a.html}}


\bibitem[\protect\citeauthoryear{Bonawitz, Eichner, Grieskamp, Huba, Ingerman,
  Ivanov, Kiddon, Konečný, Mazzocchi, McMahan, Overveldt, Petrou, Ramage, and
  Roselander}{Bonawitz et~al\mbox{.}}{2019}]%
        {federated2}
\bibfield{author}{\bibinfo{person}{K.~A. Bonawitz}, \bibinfo{person}{Hubert
  Eichner}, \bibinfo{person}{Wolfgang Grieskamp}, \bibinfo{person}{Dzmitry
  Huba}, \bibinfo{person}{Alex Ingerman}, \bibinfo{person}{Vladimir Ivanov},
  \bibinfo{person}{Chloé~M Kiddon}, \bibinfo{person}{Jakub Konečný},
  \bibinfo{person}{Stefano Mazzocchi}, \bibinfo{person}{Brendan McMahan},
  \bibinfo{person}{Timon~Van Overveldt}, \bibinfo{person}{David Petrou},
  \bibinfo{person}{Daniel Ramage}, {and} \bibinfo{person}{Jason Roselander}.}
  \bibinfo{year}{2019}\natexlab{}.
\newblock \showarticletitle{Towards Federated Learning at Scale: System
  Design}. In \bibinfo{booktitle}{{\em SysML 2019}}.
\newblock
\showURL{%
\url{https://arxiv.org/abs/1902.01046}}
\newblock
\shownote{To appear.}


\bibitem[\protect\citeauthoryear{{Brendan McMahan, Ramesh Raskar, Otkrist
  Gupta, Praneeth Vepakomma, Hassan Takabi, Jakub Konečný}}{{Brendan McMahan,
  Ramesh Raskar, Otkrist Gupta, Praneeth Vepakomma, Hassan Takabi, Jakub
  Konečný}}{2019}]%
        {CVPR_tut}
\bibfield{author}{\bibinfo{person}{{Brendan McMahan, Ramesh Raskar, Otkrist
  Gupta, Praneeth Vepakomma, Hassan Takabi, Jakub Konečný}}.}
  \bibinfo{year}{2019}\natexlab{}.
\newblock \bibinfo{title}{CVPR Tutorial On Distributed Private Machine Learning
  for Computer Vision: Federated Learning, Split Learning and Beyond}.
\newblock \bibinfo{howpublished}{\url{https://nopeekcvpr.github.io}}.
  (\bibinfo{year}{2019}).
\newblock


\bibitem[\protect\citeauthoryear{Brown, Mann, Ryder, Subbiah, Kaplan, Dhariwal,
  Neelakantan, Shyam, Sastry, Askell, Agarwal, Herbert-Voss, Krueger, Henighan,
  Child, Ramesh, Ziegler, Wu, Winter, Hesse, Chen, Sigler, Litwin, Gray, Chess,
  Clark, Berner, McCandlish, Radford, Sutskever, and Amodei}{Brown
  et~al\mbox{.}}{2020}]%
        {gpt3}
\bibfield{author}{\bibinfo{person}{Tom~B. Brown}, \bibinfo{person}{Benjamin
  Mann}, \bibinfo{person}{Nick Ryder}, \bibinfo{person}{Melanie Subbiah},
  \bibinfo{person}{Jared Kaplan}, \bibinfo{person}{Prafulla Dhariwal},
  \bibinfo{person}{Arvind Neelakantan}, \bibinfo{person}{Pranav Shyam},
  \bibinfo{person}{Girish Sastry}, \bibinfo{person}{Amanda Askell},
  \bibinfo{person}{Sandhini Agarwal}, \bibinfo{person}{Ariel Herbert-Voss},
  \bibinfo{person}{Gretchen Krueger}, \bibinfo{person}{Tom Henighan},
  \bibinfo{person}{Rewon Child}, \bibinfo{person}{Aditya Ramesh},
  \bibinfo{person}{Daniel~M. Ziegler}, \bibinfo{person}{Jeffrey Wu},
  \bibinfo{person}{Clemens Winter}, \bibinfo{person}{Christopher Hesse},
  \bibinfo{person}{Mark Chen}, \bibinfo{person}{Eric Sigler},
  \bibinfo{person}{Mateusz Litwin}, \bibinfo{person}{Scott Gray},
  \bibinfo{person}{Benjamin Chess}, \bibinfo{person}{Jack Clark},
  \bibinfo{person}{Christopher Berner}, \bibinfo{person}{Sam McCandlish},
  \bibinfo{person}{Alec Radford}, \bibinfo{person}{Ilya Sutskever}, {and}
  \bibinfo{person}{Dario Amodei}.} \bibinfo{year}{2020}\natexlab{}.
\newblock \bibinfo{title}{Language Models are Few-Shot Learners}.
\newblock   (\bibinfo{year}{2020}).
\newblock
\showeprint[arxiv]{cs.CL/2005.14165}


\bibitem[\protect\citeauthoryear{Bulat and Tzimiropoulos}{Bulat and
  Tzimiropoulos}{2017}]%
        {bulat2017far}
\bibfield{author}{\bibinfo{person}{Adrian Bulat} {and}
  \bibinfo{person}{Georgios Tzimiropoulos}.} \bibinfo{year}{2017}\natexlab{}.
\newblock \showarticletitle{How far are we from solving the 2D \& 3D Face
  Alignment problem? (and a dataset of 230,000 3D facial landmarks)}. In
  \bibinfo{booktitle}{{\em International Conference on Computer Vision}}.
\newblock


\bibitem[\protect\citeauthoryear{Ceballos, Sharma, Mugica, Singh, Roman,
  Vepakomma, and Raskar}{Ceballos et~al\mbox{.}}{2020}]%
        {splitnn8}
\bibfield{author}{\bibinfo{person}{Iker Ceballos}, \bibinfo{person}{Vivek
  Sharma}, \bibinfo{person}{Eduardo Mugica}, \bibinfo{person}{Abhishek Singh},
  \bibinfo{person}{Alberto Roman}, \bibinfo{person}{Praneeth Vepakomma}, {and}
  \bibinfo{person}{Ramesh Raskar}.} \bibinfo{year}{2020}\natexlab{}.
\newblock \bibinfo{title}{SplitNN-driven Vertical Partitioning}.
\newblock   (\bibinfo{year}{2020}).
\newblock
\showeprint[arxiv]{cs.LG/2008.04137}


\bibitem[\protect\citeauthoryear{Coates, Ng, and Lee}{Coates
  et~al\mbox{.}}{2011}]%
        {stl10}
\bibfield{author}{\bibinfo{person}{Adam Coates}, \bibinfo{person}{Andrew Ng},
  {and} \bibinfo{person}{Honglak Lee}.} \bibinfo{year}{2011}\natexlab{}.
\newblock \showarticletitle{An Analysis of Single-Layer Networks in
  Unsupervised Feature Learning}. In \bibinfo{booktitle}{{\em Proceedings of
  the Fourteenth International Conference on Artificial Intelligence and
  Statistics}} {\em (\bibinfo{series}{Proceedings of Machine Learning
  Research})}, \bibfield{editor}{\bibinfo{person}{Geoffrey Gordon},
  \bibinfo{person}{David Dunson}, {and} \bibinfo{person}{Miroslav Dudík}}
  (Eds.), Vol.~\bibinfo{volume}{15}. \bibinfo{publisher}{PMLR},
  \bibinfo{address}{Fort Lauderdale, FL, USA}, \bibinfo{pages}{215--223}.
\newblock
\showURL{%
\url{http://proceedings.mlr.press/v15/coates11a.html}}


\bibitem[\protect\citeauthoryear{Fang, Cao, Jia, and Gong}{Fang
  et~al\mbox{.}}{2020}]%
        {fed_poison}
\bibfield{author}{\bibinfo{person}{Minghong Fang}, \bibinfo{person}{Xiaoyu
  Cao}, \bibinfo{person}{Jinyuan Jia}, {and} \bibinfo{person}{Neil Gong}.}
  \bibinfo{year}{2020}\natexlab{}.
\newblock \showarticletitle{Local Model Poisoning Attacks to Byzantine-Robust
  Federated Learning}. In \bibinfo{booktitle}{{\em 29th {USENIX} Security
  Symposium ({USENIX} Security 20)}}. \bibinfo{publisher}{{USENIX}
  Association}, \bibinfo{pages}{1605--1622}.
\newblock
\showISBNx{978-1-939133-17-5}
\showURL{%
\url{https://www.usenix.org/conference/usenixsecurity20/presentation/fang}}


\bibitem[\protect\citeauthoryear{Fredrikson, Jha, and Ristenpart}{Fredrikson
  et~al\mbox{.}}{2015}]%
        {model_inv}
\bibfield{author}{\bibinfo{person}{Matt Fredrikson}, \bibinfo{person}{Somesh
  Jha}, {and} \bibinfo{person}{Thomas Ristenpart}.}
  \bibinfo{year}{2015}\natexlab{}.
\newblock \showarticletitle{Model Inversion Attacks That Exploit Confidence
  Information and Basic Countermeasures}. In \bibinfo{booktitle}{{\em
  Proceedings of the 22nd ACM SIGSAC Conference on Computer and Communications
  Security}} {\em (\bibinfo{series}{CCS '15})}. \bibinfo{publisher}{Association
  for Computing Machinery}, \bibinfo{address}{New York, NY, USA},
  \bibinfo{pages}{1322–1333}.
\newblock
\showISBNx{9781450338325}
\showDOI{%
\url{https://doi.org/10.1145/2810103.2813677}}


\bibitem[\protect\citeauthoryear{Froelicher, Troncoso-Pastoriza, Pyrgelis, Sav,
  Sousa, Bossuat, and Hubaux}{Froelicher et~al\mbox{.}}{2020}]%
        {fed_def}
\bibfield{author}{\bibinfo{person}{David Froelicher}, \bibinfo{person}{Juan~R.
  Troncoso-Pastoriza}, \bibinfo{person}{Apostolos Pyrgelis},
  \bibinfo{person}{Sinem Sav}, \bibinfo{person}{Joao~Sa Sousa},
  \bibinfo{person}{Jean-Philippe Bossuat}, {and} \bibinfo{person}{Jean-Pierre
  Hubaux}.} \bibinfo{year}{2020}\natexlab{}.
\newblock \bibinfo{title}{Scalable Privacy-Preserving Distributed Learning}.
\newblock   (\bibinfo{year}{2020}).
\newblock
\showeprint[arxiv]{cs.CR/2005.09532}


\bibitem[\protect\citeauthoryear{Fung, Yoon, and Beschastnikh}{Fung
  et~al\mbox{.}}{2020}]%
        {fed_poison_defense}
\bibfield{author}{\bibinfo{person}{Clement Fung}, \bibinfo{person}{Chris J.~M.
  Yoon}, {and} \bibinfo{person}{Ivan Beschastnikh}.}
  \bibinfo{year}{2020}\natexlab{}.
\newblock \showarticletitle{The Limitations of Federated Learning in Sybil
  Settings}. In \bibinfo{booktitle}{{\em 23rd International Symposium on
  Research in Attacks, Intrusions and Defenses ({RAID} 2020)}}.
  \bibinfo{publisher}{{USENIX} Association}, \bibinfo{address}{San Sebastian},
  \bibinfo{pages}{301--316}.
\newblock
\showISBNx{978-1-939133-18-2}
\showURL{%
\url{https://www.usenix.org/conference/raid2020/presentation/fung}}


\bibitem[\protect\citeauthoryear{Ganju, Wang, Yang, Gunter, and Borisov}{Ganju
  et~al\mbox{.}}{2018}]%
        {ccs18_inference}
\bibfield{author}{\bibinfo{person}{Karan Ganju}, \bibinfo{person}{Qi Wang},
  \bibinfo{person}{Wei Yang}, \bibinfo{person}{Carl~A. Gunter}, {and}
  \bibinfo{person}{Nikita Borisov}.} \bibinfo{year}{2018}\natexlab{}.
\newblock \showarticletitle{Property Inference Attacks on Fully Connected
  Neural Networks Using Permutation Invariant Representations}. In
  \bibinfo{booktitle}{{\em Proceedings of the 2018 ACM SIGSAC Conference on
  Computer and Communications Security}} {\em (\bibinfo{series}{CCS '18})}.
  \bibinfo{publisher}{Association for Computing Machinery},
  \bibinfo{address}{New York, NY, USA}, \bibinfo{pages}{619–633}.
\newblock
\showISBNx{9781450356930}
\showDOI{%
\url{https://doi.org/10.1145/3243734.3243834}}


\bibitem[\protect\citeauthoryear{{Gao}, {Kim}, {Abuadbba}, {Kim}, {Thapa},
  {Kim}, {Camtep}, {Kim}, and {Nepal}}{{Gao} et~al\mbox{.}}{2020}]%
        {split_iot}
\bibfield{author}{\bibinfo{person}{Y. {Gao}}, \bibinfo{person}{M. {Kim}},
  \bibinfo{person}{S. {Abuadbba}}, \bibinfo{person}{Y. {Kim}},
  \bibinfo{person}{C. {Thapa}}, \bibinfo{person}{K. {Kim}},
  \bibinfo{person}{S.~A. {Camtep}}, \bibinfo{person}{H. {Kim}}, {and}
  \bibinfo{person}{S. {Nepal}}.} \bibinfo{year}{2020}\natexlab{}.
\newblock \showarticletitle{End-to-End Evaluation of Federated Learning and
  Split Learning for Internet of Things}. In \bibinfo{booktitle}{{\em 2020
  International Symposium on Reliable Distributed Systems (SRDS)}}.
  \bibinfo{pages}{91--100}.
\newblock
\showDOI{%
\url{https://doi.org/10.1109/SRDS51746.2020.00017}}


\bibitem[\protect\citeauthoryear{Goodfellow, Pouget-Abadie, Mirza, Xu,
  Warde-Farley, Ozair, Courville, and Bengio}{Goodfellow et~al\mbox{.}}{2014}]%
        {gan}
\bibfield{author}{\bibinfo{person}{Ian Goodfellow}, \bibinfo{person}{Jean
  Pouget-Abadie}, \bibinfo{person}{Mehdi Mirza}, \bibinfo{person}{Bing Xu},
  \bibinfo{person}{David Warde-Farley}, \bibinfo{person}{Sherjil Ozair},
  \bibinfo{person}{Aaron Courville}, {and} \bibinfo{person}{Yoshua Bengio}.}
  \bibinfo{year}{2014}\natexlab{}.
\newblock \showarticletitle{Generative Adversarial Nets}. In
  \bibinfo{booktitle}{{\em Advances in Neural Information Processing Systems}},
  \bibfield{editor}{\bibinfo{person}{Z.~Ghahramani},
  \bibinfo{person}{M.~Welling}, \bibinfo{person}{C.~Cortes},
  \bibinfo{person}{N.~Lawrence}, {and} \bibinfo{person}{K.~Q. Weinberger}}
  (Eds.), Vol.~\bibinfo{volume}{27}. \bibinfo{publisher}{Curran Associates,
  Inc.}, \bibinfo{pages}{2672--2680}.
\newblock


\bibitem[\protect\citeauthoryear{Gulrajani, Ahmed, Arjovsky, Dumoulin, and
  Courville}{Gulrajani et~al\mbox{.}}{2017}]%
        {wgan}
\bibfield{author}{\bibinfo{person}{Ishaan Gulrajani}, \bibinfo{person}{Faruk
  Ahmed}, \bibinfo{person}{Martin Arjovsky}, \bibinfo{person}{Vincent
  Dumoulin}, {and} \bibinfo{person}{Aaron~C Courville}.}
  \bibinfo{year}{2017}\natexlab{}.
\newblock \showarticletitle{Improved Training of Wasserstein GANs}. In
  \bibinfo{booktitle}{{\em Advances in Neural Information Processing Systems}},
  \bibfield{editor}{\bibinfo{person}{I.~Guyon}, \bibinfo{person}{U.~V.
  Luxburg}, \bibinfo{person}{S.~Bengio}, \bibinfo{person}{H.~Wallach},
  \bibinfo{person}{R.~Fergus}, \bibinfo{person}{S.~Vishwanathan}, {and}
  \bibinfo{person}{R.~Garnett}} (Eds.), Vol.~\bibinfo{volume}{30}.
  \bibinfo{publisher}{Curran Associates, Inc.}, \bibinfo{pages}{5767--5777}.
\newblock


\bibitem[\protect\citeauthoryear{Gupta and Raskar}{Gupta and Raskar}{2018}]%
        {splitnn}
\bibfield{author}{\bibinfo{person}{Otkrist Gupta} {and} \bibinfo{person}{Ramesh
  Raskar}.} \bibinfo{year}{2018}\natexlab{}.
\newblock \showarticletitle{Distributed learning of deep neural network over
  multiple agents}.
\newblock \bibinfo{journal}{{\em Journal of Network and Computer
  Applications\/}}  \bibinfo{volume}{116} (\bibinfo{year}{2018}),
  \bibinfo{pages}{1 -- 8}.
\newblock
\showISSN{1084-8045}
\showDOI{%
\url{https://doi.org/10.1016/j.jnca.2018.05.003}}


\bibitem[\protect\citeauthoryear{Gutman, Codella, Celebi, Helba, Marchetti,
  Mishra, and Halpern}{Gutman et~al\mbox{.}}{2016}]%
        {isic}
\bibfield{author}{\bibinfo{person}{David Gutman}, \bibinfo{person}{Noel C.~F.
  Codella}, \bibinfo{person}{M.~Emre Celebi}, \bibinfo{person}{Brian Helba},
  \bibinfo{person}{Michael~A. Marchetti}, \bibinfo{person}{Nabin~K. Mishra},
  {and} \bibinfo{person}{Allan Halpern}.} \bibinfo{year}{2016}\natexlab{}.
\newblock \showarticletitle{Skin Lesion Analysis toward Melanoma Detection: {A}
  Challenge at the International Symposium on Biomedical Imaging {(ISBI)} 2016,
  hosted by the International Skin Imaging Collaboration {(ISIC)}}.
\newblock \bibinfo{journal}{{\em CoRR\/}}  \bibinfo{volume}{abs/1605.01397}
  (\bibinfo{year}{2016}).
\newblock
\showeprint[arxiv]{1605.01397}
\showURL{%
\url{http://arxiv.org/abs/1605.01397}}


\bibitem[\protect\citeauthoryear{{Hao}, {Li}, {Luo}, {Xu}, {Yang}, and
  {Liu}}{{Hao} et~al\mbox{.}}{2020}]%
        {fed_def2}
\bibfield{author}{\bibinfo{person}{M. {Hao}}, \bibinfo{person}{H. {Li}},
  \bibinfo{person}{X. {Luo}}, \bibinfo{person}{G. {Xu}}, \bibinfo{person}{H.
  {Yang}}, {and} \bibinfo{person}{S. {Liu}}.} \bibinfo{year}{2020}\natexlab{}.
\newblock \showarticletitle{Efficient and Privacy-Enhanced Federated Learning
  for Industrial Artificial Intelligence}.
\newblock \bibinfo{journal}{{\em IEEE Transactions on Industrial
  Informatics\/}} \bibinfo{volume}{16}, \bibinfo{number}{10}
  (\bibinfo{year}{2020}), \bibinfo{pages}{6532--6542}.
\newblock
\showDOI{%
\url{https://doi.org/10.1109/TII.2019.2945367}}


\bibitem[\protect\citeauthoryear{{He}, {Zhang}, {Ren}, and {Sun}}{{He}
  et~al\mbox{.}}{2016}]%
        {resnet}
\bibfield{author}{\bibinfo{person}{K. {He}}, \bibinfo{person}{X. {Zhang}},
  \bibinfo{person}{S. {Ren}}, {and} \bibinfo{person}{J. {Sun}}.}
  \bibinfo{year}{2016}\natexlab{}.
\newblock \showarticletitle{Deep Residual Learning for Image Recognition}. In
  \bibinfo{booktitle}{{\em 2016 IEEE Conference on Computer Vision and Pattern
  Recognition (CVPR)}}. \bibinfo{pages}{770--778}.
\newblock
\showDOI{%
\url{https://doi.org/10.1109/CVPR.2016.90}}


\bibitem[\protect\citeauthoryear{He, Zhang, and Lee}{He et~al\mbox{.}}{2019}]%
        {model_inversion}
\bibfield{author}{\bibinfo{person}{Zecheng He}, \bibinfo{person}{Tianwei
  Zhang}, {and} \bibinfo{person}{Ruby~B. Lee}.}
  \bibinfo{year}{2019}\natexlab{}.
\newblock \showarticletitle{Model Inversion Attacks against Collaborative
  Inference}. In \bibinfo{booktitle}{{\em Proceedings of the 35th Annual
  Computer Security Applications Conference}} {\em (\bibinfo{series}{ACSAC
  '19})}. \bibinfo{publisher}{Association for Computing Machinery},
  \bibinfo{address}{New York, NY, USA}, \bibinfo{pages}{148–162}.
\newblock
\showISBNx{9781450376280}
\showDOI{%
\url{https://doi.org/10.1145/3359789.3359824}}


\bibitem[\protect\citeauthoryear{Hitaj, Ateniese, and Perez-Cruz}{Hitaj
  et~al\mbox{.}}{2017}]%
        {gan_attack}
\bibfield{author}{\bibinfo{person}{Briland Hitaj}, \bibinfo{person}{Giuseppe
  Ateniese}, {and} \bibinfo{person}{Fernando Perez-Cruz}.}
  \bibinfo{year}{2017}\natexlab{}.
\newblock \showarticletitle{Deep Models Under the GAN: Information Leakage from
  Collaborative Deep Learning}. In \bibinfo{booktitle}{{\em Proceedings of the
  2017 ACM SIGSAC Conference on Computer and Communications Security}} {\em
  (\bibinfo{series}{CCS '17})}. \bibinfo{publisher}{Association for Computing
  Machinery}, \bibinfo{address}{New York, NY, USA}, \bibinfo{pages}{603–618}.
\newblock
\showISBNx{9781450349468}
\showDOI{%
\url{https://doi.org/10.1145/3133956.3134012}}


\bibitem[\protect\citeauthoryear{{Jeon} and {Kim}}{{Jeon} and {Kim}}{2020}]%
        {split_para}
\bibfield{author}{\bibinfo{person}{J. {Jeon}} {and} \bibinfo{person}{J.
  {Kim}}.} \bibinfo{year}{2020}\natexlab{}.
\newblock \showarticletitle{Privacy-Sensitive Parallel Split Learning}. In
  \bibinfo{booktitle}{{\em 2020 International Conference on Information
  Networking (ICOIN)}}. \bibinfo{pages}{7--9}.
\newblock
\showDOI{%
\url{https://doi.org/10.1109/ICOIN48656.2020.9016486}}


\bibitem[\protect\citeauthoryear{{Kang}, {Xiong}, {Niyato}, {Xie}, and
  {Zhang}}{{Kang} et~al\mbox{.}}{2019}]%
        {fed_def1}
\bibfield{author}{\bibinfo{person}{J. {Kang}}, \bibinfo{person}{Z. {Xiong}},
  \bibinfo{person}{D. {Niyato}}, \bibinfo{person}{S. {Xie}}, {and}
  \bibinfo{person}{J. {Zhang}}.} \bibinfo{year}{2019}\natexlab{}.
\newblock \showarticletitle{Incentive Mechanism for Reliable Federated
  Learning: A Joint Optimization Approach to Combining Reputation and Contract
  Theory}.
\newblock \bibinfo{journal}{{\em IEEE Internet of Things Journal\/}}
  \bibinfo{volume}{6}, \bibinfo{number}{6} (\bibinfo{year}{2019}),
  \bibinfo{pages}{10700--10714}.
\newblock
\showDOI{%
\url{https://doi.org/10.1109/JIOT.2019.2940820}}


\bibitem[\protect\citeauthoryear{Kim, Shin, Yu, Lee, and Lee}{Kim
  et~al\mbox{.}}{2020}]%
        {splitmc}
\bibfield{author}{\bibinfo{person}{J. Kim}, \bibinfo{person}{Sungho Shin},
  \bibinfo{person}{Yeonguk Yu}, \bibinfo{person}{Junseok Lee}, {and}
  \bibinfo{person}{Kyoobin Lee}.} \bibinfo{year}{2020}\natexlab{}.
\newblock \showarticletitle{Multiple Classification with Split Learning}.
\newblock \bibinfo{journal}{{\em ArXiv\/}}  \bibinfo{volume}{abs/2008.09874}
  (\bibinfo{year}{2020}).
\newblock


\bibitem[\protect\citeauthoryear{Koda, Park, Bennis, Yamamoto, Nishio, and
  Morikura}{Koda et~al\mbox{.}}{2019}]%
        {split_wave}
\bibfield{author}{\bibinfo{person}{Yusuke Koda}, \bibinfo{person}{Jihong Park},
  \bibinfo{person}{Mehdi Bennis}, \bibinfo{person}{Koji Yamamoto},
  \bibinfo{person}{Takayuki Nishio}, {and} \bibinfo{person}{Masahiro
  Morikura}.} \bibinfo{year}{2019}\natexlab{}.
\newblock \showarticletitle{One Pixel Image and RF Signal Based Split Learning
  for MmWave Received Power Prediction} {\em (\bibinfo{series}{CoNEXT '19
  Companion})}. \bibinfo{publisher}{Association for Computing Machinery},
  \bibinfo{address}{New York, NY, USA}, \bibinfo{pages}{54–56}.
\newblock
\showISBNx{9781450370066}
\showDOI{%
\url{https://doi.org/10.1145/3360468.3368176}}


\bibitem[\protect\citeauthoryear{Konečný, McMahan, Ramage, and
  Richtárik}{Konečný et~al\mbox{.}}{2016}]%
        {federated0}
\bibfield{author}{\bibinfo{person}{Jakub Konečný},
  \bibinfo{person}{H.~Brendan McMahan}, \bibinfo{person}{Daniel Ramage}, {and}
  \bibinfo{person}{Peter Richtárik}.} \bibinfo{year}{2016}\natexlab{}.
\newblock \bibinfo{title}{Federated Optimization: Distributed Machine Learning
  for On-Device Intelligence}.
\newblock   (\bibinfo{year}{2016}).
\newblock
\showeprint[arxiv]{cs.LG/1610.02527}


\bibitem[\protect\citeauthoryear{Konečný, McMahan, Yu, Richtárik, Suresh,
  and Bacon}{Konečný et~al\mbox{.}}{2017}]%
        {federated1}
\bibfield{author}{\bibinfo{person}{Jakub Konečný},
  \bibinfo{person}{H.~Brendan McMahan}, \bibinfo{person}{Felix~X. Yu},
  \bibinfo{person}{Peter Richtárik}, \bibinfo{person}{Ananda~Theertha Suresh},
  {and} \bibinfo{person}{Dave Bacon}.} \bibinfo{year}{2017}\natexlab{}.
\newblock \bibinfo{title}{Federated Learning: Strategies for Improving
  Communication Efficiency}.
\newblock   (\bibinfo{year}{2017}).
\newblock
\showeprint[arxiv]{cs.LG/1610.05492}


\bibitem[\protect\citeauthoryear{Lake, Salakhutdinov, and Tenenbaum}{Lake
  et~al\mbox{.}}{2015}]%
        {omni}
\bibfield{author}{\bibinfo{person}{Brenden~M. Lake}, \bibinfo{person}{Ruslan
  Salakhutdinov}, {and} \bibinfo{person}{Joshua~B. Tenenbaum}.}
  \bibinfo{year}{2015}\natexlab{}.
\newblock \showarticletitle{Human-level concept learning through probabilistic
  program induction}.
\newblock \bibinfo{journal}{{\em Science\/}} \bibinfo{volume}{350},
  \bibinfo{number}{6266} (\bibinfo{year}{2015}), \bibinfo{pages}{1332--1338}.
\newblock
\showISSN{0036-8075}
\showDOI{%
\url{https://doi.org/10.1126/science.aab3050}}
\showeprint{https://science.sciencemag.org/content/350/6266/1332.full.pdf}


\bibitem[\protect\citeauthoryear{{Langer}, {He}, {Rahayu}, and {Xue}}{{Langer}
  et~al\mbox{.}}{2020}]%
        {dist}
\bibfield{author}{\bibinfo{person}{M. {Langer}}, \bibinfo{person}{Z. {He}},
  \bibinfo{person}{W. {Rahayu}}, {and} \bibinfo{person}{Y. {Xue}}.}
  \bibinfo{year}{2020}\natexlab{}.
\newblock \showarticletitle{Distributed Training of Deep Learning Models: A
  Taxonomic Perspective}.
\newblock \bibinfo{journal}{{\em IEEE Transactions on Parallel and Distributed
  Systems\/}} \bibinfo{volume}{31}, \bibinfo{number}{12}
  (\bibinfo{year}{2020}), \bibinfo{pages}{2802--2818}.
\newblock
\showDOI{%
\url{https://doi.org/10.1109/TPDS.2020.3003307}}


\bibitem[\protect\citeauthoryear{Lim, Ng, Xiong, Niyato, Leung, Miao, and
  Yang}{Lim et~al\mbox{.}}{2020}]%
        {lim2020incentive}
\bibfield{author}{\bibinfo{person}{Wei Yang~Bryan Lim},
  \bibinfo{person}{Jer~Shyuan Ng}, \bibinfo{person}{Zehui Xiong},
  \bibinfo{person}{Dusit Niyato}, \bibinfo{person}{Cyril Leung},
  \bibinfo{person}{Chunyan Miao}, {and} \bibinfo{person}{Qiang Yang}.}
  \bibinfo{year}{2020}\natexlab{}.
\newblock \bibinfo{title}{Incentive Mechanism Design for Resource Sharing in
  Collaborative Edge Learning}.
\newblock   (\bibinfo{year}{2020}).
\newblock
\showeprint[arxiv]{cs.NI/2006.00511}


\bibitem[\protect\citeauthoryear{Liu, Luo, Wang, and Tang}{Liu
  et~al\mbox{.}}{2015}]%
        {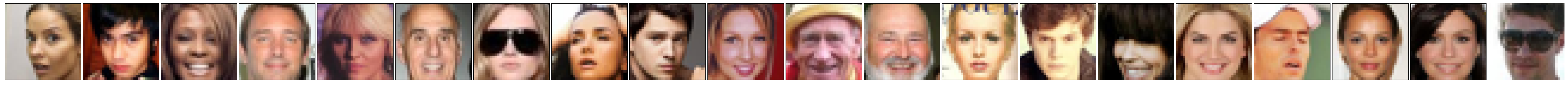}
\bibfield{author}{\bibinfo{person}{Ziwei Liu}, \bibinfo{person}{Ping Luo},
  \bibinfo{person}{Xiaogang Wang}, {and} \bibinfo{person}{Xiaoou Tang}.}
  \bibinfo{year}{2015}\natexlab{}.
\newblock \showarticletitle{Deep Learning Face Attributes in the Wild}. In
  \bibinfo{booktitle}{{\em Proceedings of International Conference on Computer
  Vision (ICCV)}}.
\newblock


\bibitem[\protect\citeauthoryear{Palanisamy, Khimani, Moti, and
  Chatzopoulos}{Palanisamy et~al\mbox{.}}{2020}]%
        {spliteasy}
\bibfield{author}{\bibinfo{person}{Kamalesh Palanisamy}, \bibinfo{person}{Vivek
  Khimani}, \bibinfo{person}{Moin~Hussain Moti}, {and} \bibinfo{person}{D.
  Chatzopoulos}.} \bibinfo{year}{2020}\natexlab{}.
\newblock \showarticletitle{SplitEasy: A Practical Approach for Training ML
  models on Mobile Devices in a split second}.
\newblock \bibinfo{journal}{{\em ArXiv\/}}  \bibinfo{volume}{abs/2011.04232}
  (\bibinfo{year}{2020}).
\newblock


\bibitem[\protect\citeauthoryear{Poirot, Vepakomma, Chang, Kalpathy-Cramer,
  Gupta, and Raskar}{Poirot et~al\mbox{.}}{2019}]%
        {splitnn3}
\bibfield{author}{\bibinfo{person}{Maarten~G. Poirot},
  \bibinfo{person}{Praneeth Vepakomma}, \bibinfo{person}{Ken Chang},
  \bibinfo{person}{Jayashree Kalpathy-Cramer}, \bibinfo{person}{Rajiv Gupta},
  {and} \bibinfo{person}{Ramesh Raskar}.} \bibinfo{year}{2019}\natexlab{}.
\newblock \bibinfo{title}{Split Learning for collaborative deep learning in
  healthcare}.
\newblock   (\bibinfo{year}{2019}).
\newblock
\showeprint[arxiv]{cs.LG/1912.12115}


\bibitem[\protect\citeauthoryear{Radford, Metz, and Chintala}{Radford
  et~al\mbox{.}}{2016}]%
        {DCGAN}
\bibfield{author}{\bibinfo{person}{Alec Radford}, \bibinfo{person}{Luke Metz},
  {and} \bibinfo{person}{Soumith Chintala}.} \bibinfo{year}{2016}\natexlab{}.
\newblock \showarticletitle{Unsupervised Representation Learning with Deep
  Convolutional Generative Adversarial Networks}. In \bibinfo{booktitle}{{\em
  4th International Conference on Learning Representations, {ICLR} 2016, San
  Juan, Puerto Rico, May 2-4, 2016, Conference Track Proceedings}},
  \bibfield{editor}{\bibinfo{person}{Yoshua Bengio} {and} \bibinfo{person}{Yann
  LeCun}} (Eds.).
\newblock
\showURL{%
\url{http://arxiv.org/abs/1511.06434}}


\bibitem[\protect\citeauthoryear{Romanini, Hall, Papadopoulos, Titcombe,
  Ismail, Cebere, Sandmann, Roehm, and Hoeh}{Romanini et~al\mbox{.}}{2021}]%
        {splitvert2}
\bibfield{author}{\bibinfo{person}{Daniele Romanini},
  \bibinfo{person}{Adam~James Hall}, \bibinfo{person}{Pavlos Papadopoulos},
  \bibinfo{person}{Tom Titcombe}, \bibinfo{person}{Abbas Ismail},
  \bibinfo{person}{Tudor Cebere}, \bibinfo{person}{Robert Sandmann},
  \bibinfo{person}{Robin Roehm}, {and} \bibinfo{person}{Michael~A. Hoeh}.}
  \bibinfo{year}{2021}\natexlab{}.
\newblock \showarticletitle{PyVertical: A Vertical Federated Learning Framework
  for Multi-headed SplitNN}. In \bibinfo{booktitle}{{\em ICLR 2021 Workshop on
  Distributed and Private Machine Learning}}.
\newblock


\bibitem[\protect\citeauthoryear{{Sattler}, {Wiedemann}, {Müller}, and
  {Samek}}{{Sattler} et~al\mbox{.}}{2020}]%
        {opt_fed1}
\bibfield{author}{\bibinfo{person}{F. {Sattler}}, \bibinfo{person}{S.
  {Wiedemann}}, \bibinfo{person}{K.~R. {Müller}}, {and} \bibinfo{person}{W.
  {Samek}}.} \bibinfo{year}{2020}\natexlab{}.
\newblock \showarticletitle{Robust and Communication-Efficient Federated
  Learning From Non-i.i.d. Data}.
\newblock \bibinfo{journal}{{\em IEEE Transactions on Neural Networks and
  Learning Systems\/}} \bibinfo{volume}{31}, \bibinfo{number}{9}
  (\bibinfo{year}{2020}), \bibinfo{pages}{3400--3413}.
\newblock
\showDOI{%
\url{https://doi.org/10.1109/TNNLS.2019.2944481}}


\bibitem[\protect\citeauthoryear{Sharma, Vepakomma, Swedish, Chang,
  Kalpathy-Cramer, and Raskar}{Sharma et~al\mbox{.}}{2019}]%
        {sharma2019expertmatcher}
\bibfield{author}{\bibinfo{person}{Vivek Sharma}, \bibinfo{person}{Praneeth
  Vepakomma}, \bibinfo{person}{Tristan Swedish}, \bibinfo{person}{Ken Chang},
  \bibinfo{person}{Jayashree Kalpathy-Cramer}, {and} \bibinfo{person}{Ramesh
  Raskar}.} \bibinfo{year}{2019}\natexlab{}.
\newblock \bibinfo{title}{ExpertMatcher: Automating ML Model Selection for
  Clients using Hidden Representations}.
\newblock   (\bibinfo{year}{2019}).
\newblock
\showeprint[arxiv]{cs.CV/1910.03731}


\bibitem[\protect\citeauthoryear{Shokri and Shmatikov}{Shokri and
  Shmatikov}{2015}]%
        {CCS15_pp}
\bibfield{author}{\bibinfo{person}{Reza Shokri} {and} \bibinfo{person}{Vitaly
  Shmatikov}.} \bibinfo{year}{2015}\natexlab{}.
\newblock \showarticletitle{Privacy-Preserving Deep Learning}. In
  \bibinfo{booktitle}{{\em Proceedings of the 22nd ACM SIGSAC Conference on
  Computer and Communications Security}} {\em (\bibinfo{series}{CCS '15})}.
  \bibinfo{publisher}{Association for Computing Machinery},
  \bibinfo{address}{New York, NY, USA}, \bibinfo{pages}{1310–1321}.
\newblock
\showISBNx{9781450338325}
\showDOI{%
\url{https://doi.org/10.1145/2810103.2813687}}


\bibitem[\protect\citeauthoryear{{Shokri}, {Stronati}, {Song}, and
  {Shmatikov}}{{Shokri} et~al\mbox{.}}{2017}]%
        {mem}
\bibfield{author}{\bibinfo{person}{R. {Shokri}}, \bibinfo{person}{M.
  {Stronati}}, \bibinfo{person}{C. {Song}}, {and} \bibinfo{person}{V.
  {Shmatikov}}.} \bibinfo{year}{2017}\natexlab{}.
\newblock \showarticletitle{Membership Inference Attacks Against Machine
  Learning Models}. In \bibinfo{booktitle}{{\em 2017 IEEE Symposium on Security
  and Privacy (SP)}}. \bibinfo{pages}{3--18}.
\newblock
\showDOI{%
\url{https://doi.org/10.1109/SP.2017.41}}


\bibitem[\protect\citeauthoryear{Singh, Vepakomma, Gupta, and Raskar}{Singh
  et~al\mbox{.}}{2019}]%
        {splitnn4}
\bibfield{author}{\bibinfo{person}{Abhishek Singh}, \bibinfo{person}{Praneeth
  Vepakomma}, \bibinfo{person}{Otkrist Gupta}, {and} \bibinfo{person}{Ramesh
  Raskar}.} \bibinfo{year}{2019}\natexlab{}.
\newblock \bibinfo{title}{Detailed comparison of communication efficiency of
  split learning and federated learning}.
\newblock   (\bibinfo{year}{2019}).
\newblock
\showeprint[arxiv]{cs.LG/1909.09145}


\bibitem[\protect\citeauthoryear{Szekely, Rizzo, and Bakirov}{Szekely
  et~al\mbox{.}}{2008}]%
        {distcorr}
\bibfield{author}{\bibinfo{person}{Gabor Szekely}, \bibinfo{person}{Maria
  Rizzo}, {and} \bibinfo{person}{Nail Bakirov}.}
  \bibinfo{year}{2008}\natexlab{}.
\newblock \showarticletitle{Measuring and Testing Dependence by Correlation of
  Distances}.
\newblock \bibinfo{journal}{{\em The Annals of Statistics\/}}
  \bibinfo{volume}{35} (\bibinfo{date}{04} \bibinfo{year}{2008}).
\newblock
\showDOI{%
\url{https://doi.org/10.1214/009053607000000505}}


\bibitem[\protect\citeauthoryear{Thapa, Chamikara, and Camtepe}{Thapa
  et~al\mbox{.}}{2020a}]%
        {splitfed}
\bibfield{author}{\bibinfo{person}{Chandra Thapa}, \bibinfo{person}{M.~A.~P.
  Chamikara}, {and} \bibinfo{person}{Seyit Camtepe}.}
  \bibinfo{year}{2020}\natexlab{a}.
\newblock \bibinfo{title}{SplitFed: When Federated Learning Meets Split
  Learning}.
\newblock   (\bibinfo{year}{2020}).
\newblock
\showeprint[arxiv]{cs.LG/2004.12088}


\bibitem[\protect\citeauthoryear{Thapa, Chamikara, and Camtepe}{Thapa
  et~al\mbox{.}}{2020b}]%
        {thapa2020advancements}
\bibfield{author}{\bibinfo{person}{Chandra Thapa}, \bibinfo{person}{M.~A.~P.
  Chamikara}, {and} \bibinfo{person}{Seyit~A. Camtepe}.}
  \bibinfo{year}{2020}\natexlab{b}.
\newblock \bibinfo{title}{Advancements of federated learning towards privacy
  preservation: from federated learning to split learning}.
\newblock   (\bibinfo{year}{2020}).
\newblock
\showeprint[arxiv]{cs.LG/2011.14818}


\bibitem[\protect\citeauthoryear{Tschandl, Rosendahl, and Kittler}{Tschandl
  et~al\mbox{.}}{2018}]%
        {ham}
\bibfield{author}{\bibinfo{person}{Philipp Tschandl}, \bibinfo{person}{Cliff
  Rosendahl}, {and} \bibinfo{person}{Harald Kittler}.}
  \bibinfo{year}{2018}\natexlab{}.
\newblock \showarticletitle{The HAM10000 dataset, a large collection of
  multi-source dermatoscopic images of common pigmented skin lesions}.
\newblock \bibinfo{journal}{{\em Scientific Data\/}} \bibinfo{volume}{5},
  \bibinfo{number}{1} (\bibinfo{year}{2018}), \bibinfo{pages}{180161}.
\newblock
\showISBNx{2052-4463}


\bibitem[\protect\citeauthoryear{Turina, Zhang, Esposito, and Matta}{Turina
  et~al\mbox{.}}{2020}]%
        {split_fed2}
\bibfield{author}{\bibinfo{person}{Valeria Turina}, \bibinfo{person}{Zongshun
  Zhang}, \bibinfo{person}{Flavio Esposito}, {and} \bibinfo{person}{Ibrahim
  Matta}.} \bibinfo{year}{2020}\natexlab{}.
\newblock \showarticletitle{Combining Split and Federated Architectures for
  Efficiency and Privacy in Deep Learning}. In \bibinfo{booktitle}{{\em
  Proceedings of the 16th International Conference on Emerging Networking
  EXperiments and Technologies}} {\em (\bibinfo{series}{CoNEXT '20})}.
  \bibinfo{publisher}{Association for Computing Machinery},
  \bibinfo{address}{New York, NY, USA}, \bibinfo{pages}{562–563}.
\newblock
\showISBNx{9781450379489}
\showDOI{%
\url{https://doi.org/10.1145/3386367.3431678}}


\bibitem[\protect\citeauthoryear{Vepakomma, Gupta, Dubey, and Raskar}{Vepakomma
  et~al\mbox{.}}{2019}]%
        {splitnnsec}
\bibfield{author}{\bibinfo{person}{Praneeth Vepakomma},
  \bibinfo{person}{Otkrist Gupta}, \bibinfo{person}{Abhimanyu Dubey}, {and}
  \bibinfo{person}{Ramesh Raskar}.} \bibinfo{year}{2019}\natexlab{}.
\newblock \showarticletitle{Reducing leakage in distributed deep learning for
  sensitive health data}.
\newblock  (\bibinfo{date}{05} \bibinfo{year}{2019}).
\newblock


\bibitem[\protect\citeauthoryear{Vepakomma, Gupta, Swedish, and
  Raskar}{Vepakomma et~al\mbox{.}}{2018a}]%
        {splitnn2}
\bibfield{author}{\bibinfo{person}{Praneeth Vepakomma},
  \bibinfo{person}{Otkrist Gupta}, \bibinfo{person}{Tristan Swedish}, {and}
  \bibinfo{person}{Ramesh Raskar}.} \bibinfo{year}{2018}\natexlab{a}.
\newblock \bibinfo{title}{Split learning for health: Distributed deep learning
  without sharing raw patient data}.
\newblock   (\bibinfo{year}{2018}).
\newblock
\showeprint[arxiv]{cs.LG/1812.00564}


\bibitem[\protect\citeauthoryear{Vepakomma, Swedish, Raskar, Gupta, and
  Dubey}{Vepakomma et~al\mbox{.}}{2018b}]%
        {splitnn_selfsurvey}
\bibfield{author}{\bibinfo{person}{Praneeth Vepakomma},
  \bibinfo{person}{Tristan Swedish}, \bibinfo{person}{Ramesh Raskar},
  \bibinfo{person}{Otkrist Gupta}, {and} \bibinfo{person}{Abhimanyu Dubey}.}
  \bibinfo{year}{2018}\natexlab{b}.
\newblock \bibinfo{title}{No Peek: A Survey of private distributed deep
  learning}.
\newblock   (\bibinfo{year}{2018}).
\newblock
\showeprint[arxiv]{cs.LG/1812.03288}


\bibitem[\protect\citeauthoryear{Vepakomma, Swedish, Raskar, Gupta, and
  Dubey}{Vepakomma et~al\mbox{.}}{2018c}]%
        {splitnnsecex}
\bibfield{author}{\bibinfo{person}{Praneeth Vepakomma},
  \bibinfo{person}{Tristan Swedish}, \bibinfo{person}{Ramesh Raskar},
  \bibinfo{person}{Otkrist Gupta}, {and} \bibinfo{person}{Abhimanyu Dubey}.}
  \bibinfo{year}{2018}\natexlab{c}.
\newblock \bibinfo{title}{No Peek: A Survey of private distributed deep
  learning}.
\newblock   (\bibinfo{year}{2018}).
\newblock
\showeprint[arxiv]{cs.LG/1812.03288}


\bibitem[\protect\citeauthoryear{{Wang}, {Wei}, and {Zhou}}{{Wang}
  et~al\mbox{.}}{2020}]%
        {opt_fed0}
\bibfield{author}{\bibinfo{person}{C. {Wang}}, \bibinfo{person}{X. {Wei}},
  {and} \bibinfo{person}{P. {Zhou}}.} \bibinfo{year}{2020}\natexlab{}.
\newblock \showarticletitle{Optimize Scheduling of Federated Learning on
  Battery-powered Mobile Devices}. In \bibinfo{booktitle}{{\em 2020 IEEE
  International Parallel and Distributed Processing Symposium (IPDPS)}}.
  \bibinfo{pages}{212--221}.
\newblock
\showDOI{%
\url{https://doi.org/10.1109/IPDPS47924.2020.00031}}


\bibitem[\protect\citeauthoryear{Wu, Zhang, and Xu}{Wu et~al\mbox{.}}{2020}]%
        {tiny}
\bibfield{author}{\bibinfo{person}{Jiayu Wu}, \bibinfo{person}{Qixiang Zhang},
  {and} \bibinfo{person}{Guoxi Xu}.} \bibinfo{year}{2020}\natexlab{}.
\newblock \bibinfo{title}{Tiny ImageNet Challenge}.
\newblock
  \bibinfo{howpublished}{\url{http://cs231n.stanford.edu/reports/2017/pdfs/930.pdf}}.
    (\bibinfo{year}{2020}).
\newblock


\bibitem[\protect\citeauthoryear{Xiao, Rasul, and Vollgraf}{Xiao
  et~al\mbox{.}}{2017}]%
        {fashionmnist}
\bibfield{author}{\bibinfo{person}{Han Xiao}, \bibinfo{person}{Kashif Rasul},
  {and} \bibinfo{person}{Roland Vollgraf}.} \bibinfo{year}{2017}\natexlab{}.
\newblock \bibinfo{title}{Fashion-MNIST: a Novel Image Dataset for Benchmarking
  Machine Learning Algorithms}.
\newblock   (\bibinfo{year}{2017}).
\newblock
\showeprint[arxiv]{cs.LG/1708.07747}


\bibitem[\protect\citeauthoryear{Zhang and Qi}{Zhang and Qi}{2017}]%
        {utkface}
\bibfield{author}{\bibinfo{person}{Song~Yang Zhang, Zhifei} {and}
  \bibinfo{person}{Hairong Qi}.} \bibinfo{year}{2017}\natexlab{}.
\newblock \showarticletitle{Age Progression/Regression by Conditional
  Adversarial Autoencoder}. In \bibinfo{booktitle}{{\em IEEE Conference on
  Computer Vision and Pattern Recognition (CVPR)}}. IEEE.
\newblock


\bibitem[\protect\citeauthoryear{{Zhang}, {Jia}, {Pei}, {Wang}, {Li}, and
  {Song}}{{Zhang} et~al\mbox{.}}{2020}]%
        {bah}
\bibfield{author}{\bibinfo{person}{Y. {Zhang}}, \bibinfo{person}{R. {Jia}},
  \bibinfo{person}{H. {Pei}}, \bibinfo{person}{W. {Wang}}, \bibinfo{person}{B.
  {Li}}, {and} \bibinfo{person}{D. {Song}}.} \bibinfo{year}{2020}\natexlab{}.
\newblock \showarticletitle{The Secret Revealer: Generative Model-Inversion
  Attacks Against Deep Neural Networks}. In \bibinfo{booktitle}{{\em 2020
  IEEE/CVF Conference on Computer Vision and Pattern Recognition (CVPR)}}.
  \bibinfo{pages}{250--258}.
\newblock
\showDOI{%
\url{https://doi.org/10.1109/CVPR42600.2020.00033}}


\bibitem[\protect\citeauthoryear{Zhu, Liu, and Han}{Zhu et~al\mbox{.}}{2019}]%
        {DLG}
\bibfield{author}{\bibinfo{person}{Ligeng Zhu}, \bibinfo{person}{Zhijian Liu},
  {and} \bibinfo{person}{Song Han}.} \bibinfo{year}{2019}\natexlab{}.
\newblock \showarticletitle{Deep Leakage from Gradients}. In
  \bibinfo{booktitle}{{\em Advances in Neural Information Processing Systems}},
  \bibfield{editor}{\bibinfo{person}{H.~Wallach},
  \bibinfo{person}{H.~Larochelle}, \bibinfo{person}{A.~Beygelzimer},
  \bibinfo{person}{F.~d\textquotesingle Alch\'{e}-Buc},
  \bibinfo{person}{E.~Fox}, {and} \bibinfo{person}{R.~Garnett}} (Eds.),
  Vol.~\bibinfo{volume}{32}. \bibinfo{publisher}{Curran Associates, Inc.}
\newblock
\showURL{%
\url{https://proceedings.neurips.cc/paper/2019/file/60a6c4002cc7b29142def8871531281a-Paper.pdf}}


\end{thebibliography}
\section*{Appendices}
\setcounter{section}{0}
\counterwithin{table}{section}
\counterwithin{figure}{section}
\renewcommand{\thesection}{\Alph{section}}%
\section{Additional results}
\label{app:add_res_all}
\begin{figure*}[t]
	\centering
	\fboxsep=0.01mm
	\fboxrule=0mm
	\centering
	\begin{subfigure}{.9\linewidth}
		\fcolorbox{black!50}{black!50}{\includegraphics[trim = 0mm 0mm 0mm 0mm, clip, width=1\linewidth]{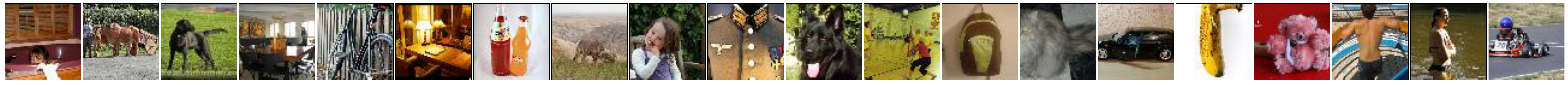}}\\
		\fcolorbox{red!50}{red!50}{\includegraphics[trim = 0mm 0mm 0mm 0mm, clip, width=1\linewidth, frame]{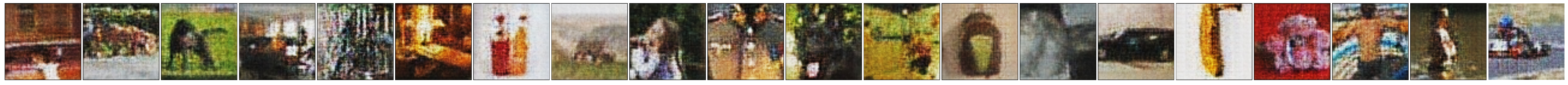}}\\
		\fcolorbox{blue!50}{blue!50}{\includegraphics[trim = 0mm 0mm 0mm 0mm, clip, width=1\linewidth, frame]{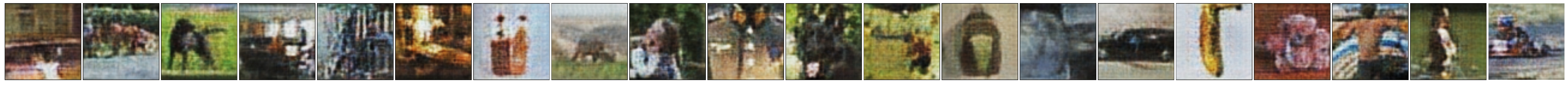}}
	\end{subfigure}\\
	
	\caption{\new{Random examples of inference of private training instances on the \textit{TinyImageNet} dataset. The first row (\ie gray frame) reports the original data, the second row (\ie red frame) depicts the attacker's reconstruction using $\xpu \myeq \textit{\textit{TinyImageNet}}$ (test) and the third row (\ie blue frame) depicts the attacker's reconstruction using $\xpu \myeq \textit{\textit{STL-10}}$. We run the attacks for $2\cdot 10^{3}$.}}
	\label{fig:tinistl20}
\end{figure*}
\begin{figure*}[t]
	\centering
	\fboxsep=0.01mm
	\fboxrule=0mm
	\centering
	\begin{subfigure}{.9\linewidth}
		\fcolorbox{black!50}{black!50}{\includegraphics[trim = 0mm 0mm 0mm 0mm, clip, width=1\linewidth]{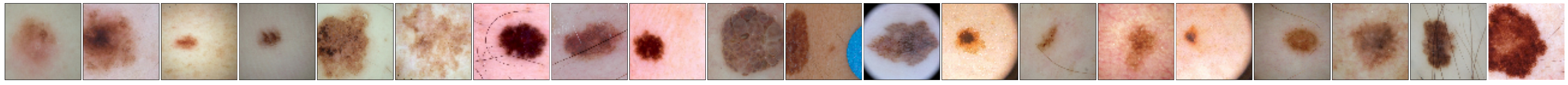}}\\
		\fcolorbox{red!50}{red!50}{\includegraphics[trim = 0mm 0mm 0mm 0mm, clip, width=1\linewidth, frame]{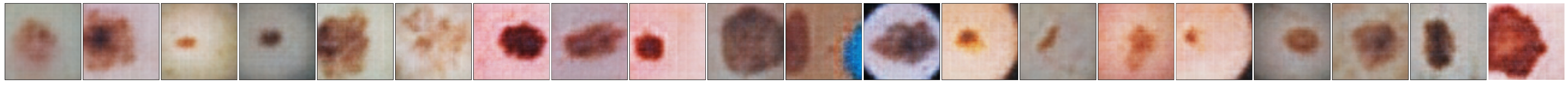}}\\
		\fcolorbox{blue!50}{blue!50}{\includegraphics[trim = 0mm 0mm 0mm 0mm, clip, width=1\linewidth, frame]{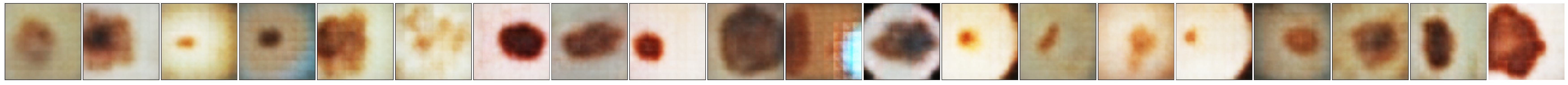}}
	\end{subfigure}\\
	
	\caption{\new{Random examples of inference of private training instances on the \textit{HAM10000} dataset. The first row (\ie gray frame) reports the original data, the second row (\ie red frame) depicts the attacker's reconstruction using $\xpu \myeq \textit{\textit{HAM10000}}$ (test) and the third row (\ie blue frame) depicts the attacker's reconstruction using $\xpu \myeq \textit{\textit{ISIC-2016}}$. We run the attacks for $2\cdot 10^{3}$.}}
	\label{fig:ham20}
\end{figure*}
\new{In this section, we include and discuss additional results.}

\subsection{On the effect of the public dataset}
\label{app:add_res}
\new{
Extending the results presented in Section~\ref{sec:xpubdiff}, we test the FSHA on other datasets.}
\new{
\paragraph{Natural images} 
Here, we test the datasets \textit{TinyImageNet} and \textit{STL-10}~\cite{stl10}. 
\textit{TinyImageNet}~\cite{tiny} is a subset of \textit{ImageNet} containing only $200$ classes of natural images. \textit{STL-10}, as \textit{TinyImageNet}, is defined over the natural domain, but it consists of only $10$ different classes (six animals and four vehicles). Note that, given the size of \textit{TinyImageNet}, the $10$ classes of \textit{STL-10} can be considered a subset of the $200$ classes of \textit{TinyImageNet}. However, there is no intersection between the images of the two sets.}
\par
\new{
Next, we test the ability of FSHA to reconstruct instances of \textit{TinyImageNet} ($\xpr$) by using the \textit{STL-10} as $\xpu$. Arguably, this attack is particularly challenging as there is a strong discrepancy between the public and private distributions. There are around $190$ unknown semantic classes of data (\ie $95\%$ of the private distribution) that the attacker has never observed. Nevertheless, as shown in Figure~\ref{fig:tinistl20}, besides altered colors and missing details, the attack converges towards suitable reconstructions of the private instances of the \textit{TinyImageNet} set, threatening clients' privacy also in this difficult setup. Again, this result suggests that the FSHA can generalize over the adopted public set and provide a representative feature space that captures unknown clients' private instances.}
\new{\paragraph{Medical images} Next, we report additional examples using dermoscopic lesion images datesets such as \textit{HAM10000}~\cite{ham} and \textit{ISIC-2016} competition dataset (task 1)~\cite{isic}.}
\par
	
\new{
	 \textit{HAM10000} is an extensive collection of multi-source dermatoscopic images of common pigmented skin lesions, containing $10015$ images collected from different populations and acquired by different modalities. \textit{ISIC-2016}, similarly to \textit{HAM10000}, collects dermatoscopic images of skin cancer, but it shows consistently less diversity in its composition and contains only $900$ images. Note that there is no intersection between the images of these two datasets.}
 \par
 
 \new{
Also in this case, we test the worst case scenario: $\xpu \myeq \textit{ISIC-2016}$ with $\xpr \myeq \textit{HAM10000}$ . Samples from the attack are reported in Figure~\ref{fig:ham20}. As in the previous case, the attack leads to the reconstruction of clients' private instances. }
\par

\new{
In the real-world scenario, the recovered images can be directly used to re-identify patients, possibly violating privacy rules.
}
\subsection{Property inference attacks}
\begin{figure}
	\includegraphics[trim = 0mm 0mm 0mm 0mm, clip, width=.7\linewidth]{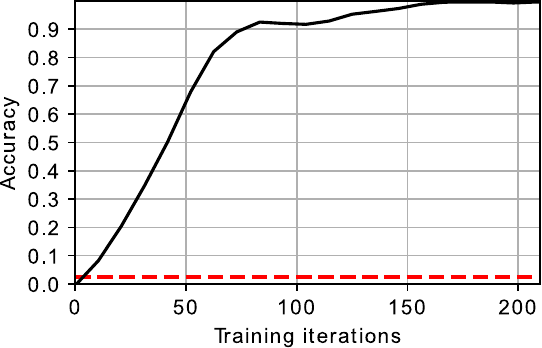}\\
	\caption{Classification accuracy during the setup phase of the FSHA \new{performed on split $4$} on the AT\&T dataset. The red, dashed line marks random guessing.}
	\label{fig:acc_id}
\end{figure}
\paragraph{Inferring categorical attributes}
\label{app:inf_cat}
The attacker can infer categorical attributes rather than binary ones by training the network $\fc$ in a multi-class classification and providing suitable labels to $\xpu$. To implement this scenario, we use the \mb{AT\&T dataset} which is composed of frontal shots of $40$ different individuals: $10$ images each. This dataset has been previously used in ~\cite{gan_attack}. Here, the server wants to identify the individuals represented on each image used during the distributed training. That is, the attacker wants to correctly assign one of the $40$ possible identities (\ie classes) to each received \textit{smashed} data.
\par

As for the previous attack, we use a single fully-connected layer to implement $\fc$ (with $40$ output units), but we train the model with a categorical cross-entropy loss function. Figure~\ref{fig:acc_id} reports the evolution of the classification accuracy during the setup phase of the attack on $\xpr$. Within a few initial iterations, the attacker reaches a perfect accuracy in classifying the images of the $40$ different individuals composing the set.
\begin{algorithm}[b]
	\footnotesize
	\KwData{number of filters: $nf$, stride $s$}
	$x = \texttt{ReLU}(x)$\;
	$x = \texttt{2D-Conv(x, nf, 3, (s,s))}$\;
	$x = \texttt{ReLU}(x)$\;
	$x = \texttt{2D-Conv(x, nf, 3, (1,1))}$\;
	\If{$s>1$}{
		$x_{in} = $\texttt{2D-Conv($x_{in}$, nf, 3, (s,s))}\;
	}
	\Return $ x_{in} + x $
	\caption{Residual Block: \texttt{resBlock}:}
	\label{algo:rb}
\end{algorithm}
\begin{table*}
	\begin{center}
		\caption{Architectures used for running the Feature-space hijacking attack.}
		\label{tab:arch}
		\resizebox{.83\textwidth}{!}{%
			\begin{tabular}{c|l|l|l|l}
				\toprule 
				\multicolumn{1}{c|}{\LARGE Split} & \multicolumn{1}{c|}{\LARGE$\f$} & \multicolumn{1}{c|}{\LARGE$\tf$} & \multicolumn{1}{c|}{\LARGE$\tfi$} & \multicolumn{1}{c}{\LARGE$D$}\\
				\midrule 
				&\texttt{2D-Conv(64, 3, (1,1), ReLU)} & \texttt{2D-Conv(64, 3, (2,2), \new{linear})} & \texttt{2D-ConvTrans(256, 3, (2,2), \new{linear})} & \texttt{2D-Conv(128, 3, (2,2), ReLU)} \\
				
				& \texttt{batch-normalization} & \texttt{2D-Conv(64, 3, (1,1), \new{linear})} & \texttt{2D-Conv(3, 3, (1,1), tanh)} & \texttt{2D-Conv(128, 3, (2,2))} \\
				
				&\texttt{ReLU} & & & \texttt{resBlock(256, 1)} \\

				\Large 1 &\texttt{maxPolling((2,2))} & & & \texttt{resBlock(256, 1)} \\
				
				&\texttt{resBlock(64, 1)} & & & \texttt{resBlock(256, 1)} \\
				
				&& & & \texttt{resBlock(256, 1)} \\
				&& & & \texttt{resBlock(256, 1)} \\
				&& & & \texttt{2D-Conv(256, 3, (2,2), ReLU)} \\
				&& & & \texttt{dense(1)} \\
				\midrule 
				
				
				&\texttt{2D-Conv(64, 3, (1,1), ReLU)} & \texttt{2D-Conv(64, 3, (2,2), \new{linear})} & \texttt{2D-ConvTrans(256, 3, (2,2), \new{linear})} & \texttt{2D-Conv(128, 3, (2,2))} \\
				
				& \texttt{batch-normalization} & \texttt{2D-Conv(128, 3, (2,2), \new{linear})} & \texttt{2D-ConvTrans(128, 3, (2,2), \new{linear})} & \texttt{resBlock(256, 1)} \\
				
				&\texttt{ReLU} & \texttt{2D-Conv(128, 3, (1,1)} & \texttt{2D-Conv(3, 3, (1,1), tanh)} & \texttt{resBlock(256, 1)} \\

				\Large 2 &\texttt{maxPolling((2,2))} & & & \texttt{resBlock(256, 1)} \\
				
				&\texttt{resBlock(64, 1)} & & & \texttt{resBlock(256, 1)} \\
				
				&\texttt{resBlock(128, 2)}& & & \texttt{resBlock(256, 1)} \\
				&& & & \texttt{2D-Conv(256, 3, (2,2), ReLU)} \\
				&& & & \texttt{dense(1)} \\
				\midrule 
				
							
			&\texttt{2D-Conv(64, 3, (1,1), ReLU)} & \texttt{2D-Conv(64, 3, (2,2), \new{linear})} & \texttt{2D-ConvTrans(256, 3, (2,2), \new{linear})} & \texttt{2D-Conv(128, 3, (2,2))} \\
			
			& \texttt{batch-normalization} & \texttt{2D-Conv(128, 3, (2,2), \new{linear})} & \texttt{2D-ConvTrans(128, 3, (2,2), \new{linear})} & \texttt{resBlock(256, 1)} \\
			
			&\texttt{ReLU} & \texttt{2D-Conv(128, 3, (1,1)} & \texttt{2D-Conv(3, 3, (1,1), tanh)} & \texttt{resBlock(256, 1)} \\

			\Large 3 &\texttt{maxPolling((2,2))} & & & \texttt{resBlock(256, 1)} \\
			
			&\texttt{resBlock(64, (1,1))} & & & \texttt{resBlock(256, 1)} \\
			
			&\texttt{resBlock(128, 2)}& & & \texttt{resBlock(256, 1)} \\
			&\texttt{resBlock(128, 1)}& & & \texttt{2D-Conv(256, 3, (2,2), ReLU)} \\
			&& & & \texttt{dense(1)} \\
			\midrule 
			
			&\texttt{2D-Conv(64, 3, (1,1), ReLU)} & \texttt{2D-Conv(64, 3, (2,2), \new{linear})} & \texttt{2D-ConvTrans(256, 3, (2,2), \new{linear})} & \texttt{2D-Conv(128, 3, (1,1))} \\
			
			& \texttt{batch-normalization} & \texttt{2D-Conv(128, 3, (2,2), \new{linear})} & \texttt{2D-ConvTrans(128, 3, (2,2), \new{linear})} & \texttt{resBlock(256, 1)} \\
			
			&\texttt{ReLU} & \texttt{2D-Conv(256, 3, (2,2), \new{linear})} & \texttt{2D-ConvTrans(3, 3, (2,2), tanh)} & \texttt{resBlock(256, 1)} \\

			\Large 4 &\texttt{maxPolling((2,2))} & \texttt{2D-Conv(256, 3, (1,1))} & & \texttt{resBlock(256, 1)} \\
			
			&\texttt{resBlock(64, 1)} & & & \texttt{resBlock(256, 1)} \\
			
			&\texttt{resBlock(128, 2)}& & & \texttt{resBlock(256, 1)} \\
			&\texttt{resBlock(128, 1)}& & & \texttt{2D-Conv(256, 3, (2,2), ReLU)} \\
			&\texttt{resBlock(256, 2)}& & & \texttt{dense(1)} \\

				\bottomrule 
			\end{tabular}	
		}
	\end{center}
\end{table*}
\section{Architectures and Experimental setups}
\label{app:arch}
The employed architectures are reported in Table~\ref{tab:arch}. For the definition of convolutional layers we use the notation: 
\begin{center}
\small \TT{(number of filters, kernel size, stride, activation function)},
\end{center}
 whereas for dense layers: 
 \begin{center}
 \small \TT{(number of nodes, activation function)}.
\end{center}
The residual block used to build the discriminator $D$ is described in Algorithm~\ref{algo:rb}.
\par

To construct the clients' network $f$, we use a standard convolutional neural network (CNN) composed of convolutional and pooling layers. The attacker's network $\tf$ outputs a tensor with the same shape of $\f$ but diverges in every other parameter. Besides being a CNN as well, $\f$ builds on different kernel sizes, kernel numbers, and activation functions; $\tf$ does not include pooling layers, but it reduces the kernel's width by a larger stride in the convolutional layers.
\par

In our experiments, we have intentionally chosen the architectures of $\f$ and $\tf$ to be different. Our aim is to be compliant with the defined threat model. However, we observed that choosing $\tf$ to be similar to $\f$ speeds up the attack procedure significantly.
\par

Table~\ref{tab:hp} reports additional hyper-parameters adopted during the attack.
\paragraph{Datasets preparation}
In our experiments, all the images on the datasets \textit{MNIST}, \textit{Fashion-MNIST}, \textit{Omniglot} and \textit{AT\&T} have been reshaped into $32\times 32 \times 3$ tensors by replicating three times the channel dimension. For the datasets \textit{CelebA}, \textit{UTKFace}, we cropped and centered the images with~\cite{bulat2017far} and reshaped them with a resolution of $64\times 64$. \textit{TinyImageNet}, \textit{STL-10}, \textit{HAM10000} and \textit{ISIC-2016} have been reshaped within a resolution of $64\times 64$.
\par

For each dataset, color intensities are scaled in the real interval $[-1,1]$.

\subsection{Client-side attack}
\label{app:csa}
To implement the client-side attack, we rely on a DCGAN-like~\cite{DCGAN} \mb{architecture} as in~\cite{gan_attack}. Specifically, the architecture for the splits $\f, \s$ and $\f^{'}$ as well as for the generator $G$ are detailed in Table~\ref{tab:arch_clienta}. As in~\cite{gan_attack}, we use a latent space of cardinality $100$ with standard, Gaussian prior.
\begin{table}
\begin{center}
	\caption{Architectures for the client-side attacks.}
	\label{tab:arch_clienta}
	\resizebox{.2\textwidth}{!}{%
		\begin{tabular}{l}
			\toprule 
			\multicolumn{1}{c}{\LARGE $\f$} \\ 
			\midrule 
			\texttt{2D-Conv(64, 5, (2,2))} \\
			\texttt{LeakyReLU}\\
			\texttt{dropout(p=0.3)} \\
			\toprule 
			\multicolumn{1}{c}{\LARGE $\s$} \\ 
			\texttt{2D-Conv(126, 5, (2,2)} \\
			\texttt{LeakyReLU}\\
			\texttt{dropout(p=0.3)} \\
			\toprule 
			\multicolumn{1}{c}{\LARGE $\f^{'}$} \\ 
			\texttt{dense(\#classes)}\\
			\texttt{sigmoid}\\
			\toprule 
			\multicolumn{1}{c}{\LARGE $G$} \\ 
			\texttt{dense(7$\cdot$7$\cdot$256)}\\
			\texttt{batch-normalization}\\
			\texttt{LeakyReLU}\\
			\texttt{2D-ConvTrans(128, 5, (1,1))}\\
			\texttt{batch-normalization}\\
						\texttt{2D-ConvTrans(128, 5, (1,1))}\\
			\texttt{batch-normalization}\\
			\texttt{LeakyReLU}\\
			\texttt{2D-ConvTrans(64, 5, (2,2))}\\
			\texttt{batch-normalization}\\
			\texttt{LeakyReLU}\\
			\texttt{2D-ConvTrans(1, 5, (2,2), tanh)}\\
			\bottomrule 
\end{tabular}	
}
\end{center}
\end{table}
%
%
%
\begin{table}
	\begin{center}
		\caption{Other hyper-parameters used during the Feature-space hijacking attack.}
		\label{tab:hp}
		\resizebox{.35\textwidth}{!}{%
			\begin{tabular}{l|r}
				\toprule 
				Optimizer $\f$ & Adam with $lr=0.00001$\\
				Optimizer $\tf$ and $\tfi$ & Adam with $lr=0.00001$\\
				Optimizer $D$ & Adam with $lr=0.0001$ \\
				& $lr=0.0005$ for split $4$ of $\f$\\
				Weight gradient penalty $D$ & 500.0\\
				\bottomrule 
			\end{tabular}	
		}
	\end{center}
\end{table}
%

\section{Evading the distance correlation metric via adversarial feature spaces}
\label{app:afsa}
\begin{figure}[t]
	\centering
	\includegraphics[trim =0mm 130mm 8mm 5mm, clip, width=.9\linewidth]{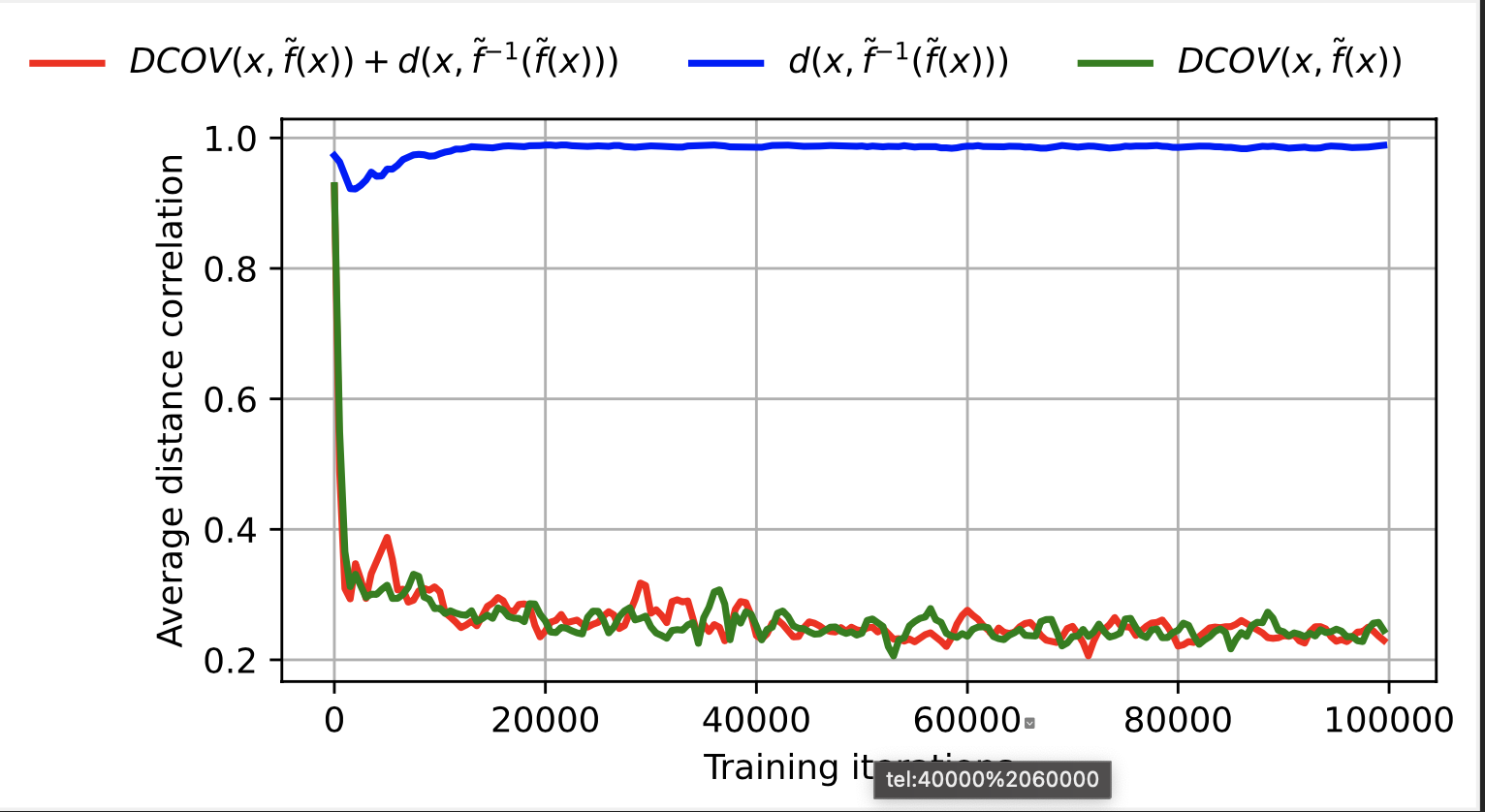}\\
	\begin{subfigure}{.4\textwidth}
		\centering
		\includegraphics[trim = 0mm 0mm 0mm 0mm, clip, width=.95\linewidth]{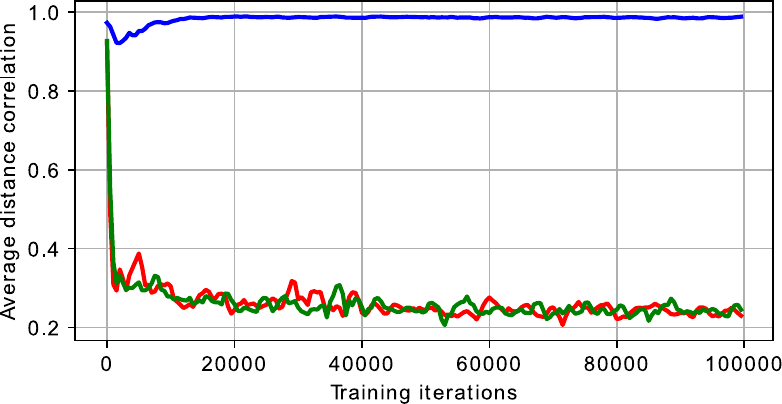}
		\caption{Distance correlation.}\label{fig:advfsdc_a}
	\end{subfigure}\\\begin{subfigure}{.4\textwidth}
		\centering
		\includegraphics[trim = 0mm 0mm 0mm 0mm, clip, width=1\linewidth]{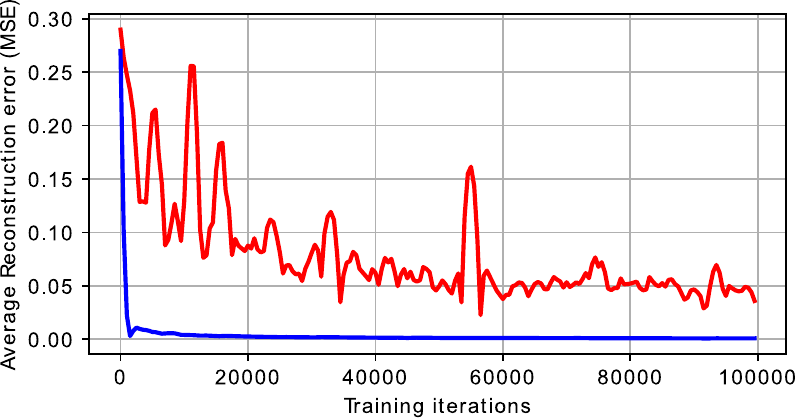}
		\caption{Reconstruction error.}\label{fig:advfsdc_b}
	\end{subfigure}
	\caption{The average distance correlation (panel (a)) and average reconstruction error (panel (b)) for the same model trained with three different losses on \textit{CelebA}.}
	\label{fig:advfsdc}
\end{figure}
Despite the proven capability of the distance correlation metrics of capturing linear as well as non-linear dependence on high-dimensional data, this can be easily evaded by highly complex mappings \mb{like those defined by deep neural networks. More formally, given an input space $X$, it is} quite simple to define a function $\f$ such that:
\begin{equation}
	\mathbb{E}_{x\sim X}[DCOR(x, \f(x))] = \epsilon_1\ \text{ , but } \ \mathbb{E}_{x\sim X}[d(x, \tfi(\f(x)))] = \epsilon_2,  
\end{equation}
where $\tfi$ is a decoder function, $d$ is a distance function defined on $X$ and $\epsilon_1$ and $ \epsilon_2$ are two constant \mb{values close to $0$. That is, the function $\f(x)$ produces an output $z$ that has minimal distance correlation with the input but that allows a} decoder network $\tfi$ to accurately recover $x$ from $z$. Intuitively, this is achieved by hiding information about $x$ in $z$ (smashed data) by allocating it in the blind spots of distance correlation metrics.
\par

In practice, such function $\f$ can be learned \mb{by tuning a neural network to minimize the following loss function:
\begin{equation}
	\loss_{\f, \tfi} = DCOR(x, \f(x)) + 	\alpha_2 \cdot d(x, \tfi(\f(x)))
	\label{eq:advdc}
\end{equation}
that is, training the network to simultaneously produce outputs that minimize their} distance correlation with the input and enable reconstruction of the input from the decoder $\tfi$. Next, we validate this idea empirically.
\par

We report the result for \textit{CelebA} and use $\f$ and $\tfi$ from the setup $4$. We use $MSE$ as $d$ and $\alpha_2=50$. We train the model for $10^{4}$ iterations. Figure~\ref{fig:advfsdc} reports the average distance correlation (Figure~\ref{fig:advfsdc_a}) and average reconstruction error (Figure~\ref{fig:advfsdc_b}) for the same model trained with three different losses; namely:
\begin{enumerate}
	\item In red, the model is trained on the adversarial loss reported in Eq.~\ref{eq:advdc}.
	\item In green, \mb{the} model is trained only to minimize distance correlation.
	\item In blue, \mb{the} model is trained only to minimize the reconstruction error (\ie auto-encoder).
\end{enumerate}
As can \mb{be noticed}, the adversarial training procedure permits to learn a pair of $\f$ and $\tfi$ such that the distance correlation is minimized (the same as we train the model only to minimize distance correlation), whereas \mb {it enables the reconstruction of the input} data.

\end{document}